\documentclass[prb, reprint, superscriptaddress]{revtex4-2}
\usepackage{amsmath,graphicx,comment,booktabs,multirow,subcaption,siunitx,float,mathtools}
\usepackage[colorlinks, allcolors=blue]{hyperref}
\usepackage{float}
\begin{document}
\title{Tunable chiral symmetry breaking in symmetric Weyl materials}
\author{Sahal Kaushik}
\email{sahal.kaushik@stonybrook.edu}
\affiliation{Department of Physics and Astronomy, Stony Brook University, Stony Brook, NY 11794, USA}
\author{Evan John Philip}
\email{ephilip@bnl.gov}
\affiliation{Department of Physics and Astronomy, Stony Brook University, Stony Brook, NY 11794, USA}
\affiliation{Computational Science Initiative, Brookhaven National Laboratory, Upton, NY 11973, USA}
\author{Jennifer Cano}
\email{jennifer.cano@stonybrook.edu}
\affiliation{Department of Physics and Astronomy, Stony Brook University, Stony Brook, NY 11794, USA}
\affiliation{Center for Computational Quantum Physics, Flatiron Institute, New York, NY 10010, USA}

\begin{abstract}
Asymmetric Weyl semimetals, which possess an inherently chiral structure, have different energies and dispersion relations for left- and right-handed fermions. They exhibit certain effects not found in symmetric Weyl semimetals, such as the quantized circular photogalvanic effect and the helical magnetic effect. In this work, we derive the conditions required for breaking chiral symmetry by applying an external field in symmetric Weyl semimetals. We explicitly demonstrate that in certain materials with the $T_d$ point group, magnetic fields along low symmetry directions break the symmetry between left- and right-handed fermions; the symmetry breaking can be tuned by changing the direction and magnitude of the magnetic field.
In some cases, we find an imbalance between the number of type I left- and right-handed Weyl cones (which is compensated by the number of type II cones of each chirality.)
\end{abstract}
\maketitle

\section{Introduction}

Dirac \cite{Young12,Wang12,Liu14,Liu14a,Steinberg14,nagaosa2014,bradlyn2016beyond,cano2019multifold,klemenz2020systematic,wieder2020strong} and Weyl \cite{Wan11,Weng15,Huang15,xu2015discovery,lv2015Nat,xu2015,lv2015PRX,2015Xiong} semimetals have linear and gapless dispersion relations, making them effective low-energy massless fermions (see Ref.~\cite{armitage2018} for a review). 
They exhibit interesting effects such as the chiral magnetic effect~\cite{2008Fukushima} and negative longitudinal  magnetoresistance~\cite{aji2012adler,2013Son,2014Burkov,2016Li, 2015Xiong,Huang2015, Zhang2016, Wang2016, Arnold2016}.

For massless fermions the chirality is equivalent to the helicity, i.e. the sign of the product of momentum and (pseudo)spin. 
Dirac semimetals have doubly degenerate bands, with fermions of both chiralities at the same point in the Brillouin zone; their band crossings are not topologically protected, but can be symmetry protected~\cite{nagaosa2014}.
In contrast, in Weyl semimetals, fermions of different chiralities are separated in momentum space.  
Consequently, Weyl semimetals exhibit some chirality-dependent effects not found in Dirac semimetals, such as a photocurrent induced by circularly polarized light~\cite{lee2017,ma2017}, and elliptically polarized terahertz emission in response to pulsed circularly polarized near infrared light \cite{gao2020chiral}.

Certain phenomena require the left- and right-handed fermions to have different energies or velocities. They include the quantized circular photogalvanic effect~\cite{dejuan2017}, the helical magnetic effect~\cite{yuta2018} and the chiral magnetic effect without an external source of chirality (i.e., a chiral chemical potential is generated because of the chirality of the crystal structure, not because of external chiral fields)~\cite{meyer2018}. These effects are possible only in materials that have chiral crystal lattices which lack symmetries that reverse spatial orientation (these include inversion, mirrors, rotoinversions and their products with time reversal). We refer to such materials as asymmetric Weyl materials, and materials that do have orientation-reversing symmetries as symmetric Weyl materials. Asymmetric Weyl materials include RhSi~\cite{chang2017} and CoSi~\cite{rao2019}.  The quantized circular photogalvanic effect has been observed in both these materials \cite{rees2020,ni2020}.

However, asymmetric Weyl materials are rare in nature compared to symmetric Weyl materials. One way to observe effects that depend on asymmetry is to break relevant symmetries by an external perturbation. 
It has recently been shown that a symmetric Weyl material can become asymmetric upon ordering magnetically if the magnetic moments break all symmetries that reverse spatial orientation \cite{ray2020}.

In this work, we investigate chiral symmetry breaking more generally.
Using the concept of true and false chirality introduced by Barron \cite{barron1986}, we derive a criterion 
to determine whether an external field or perturbation produces an asymmetric material, which depends on the symmetry of the perturbation and the space group of the crystal.
We then explicitly show that in the zincblende material InAs, applying a magnetic field in a low symmetry direction breaks all symmetries that reverse spatial orientation, and causes left- and right-handed fermions to have different velocities and energies.
This induced asymmetry allows for the observation of effects present only in asymmetric Weyl materials.
Furthermore, we show examples where the number of type I Weyl fermions \cite{soluyanov2015} of left- and right-chirality are not equal; the imbalance is compensated by the number of type II Weyl fermions of each chirality.

\section{Properties of Weyl Cones}
The chirality of a Weyl fermion is equivalent to its helicity: $\chi = \mathrm{sgn}(\vec{s}\cdot\vec{p})$ where $\vec{s}$ is the pseudospin and $\vec{p}$ is the momentum. The chirality is positive for right-handed fermions and negative for left- handed fermions.
Since Weyl cones are monopoles of Berry curvature and the total Berry charge in the Brillouin zone must be zero, there is always an equal number of left- and right-handed Weyl cones in a microscopic Hamiltonian. 

Under inversion symmetry ($P$), the pseudospin, momentum, and chirality transform as
\begin{align}
\vec{s} &\to \vec{s}\nonumber\\
\vec{p} &\to -\vec{p}\\
\chi &\to -\chi\nonumber
\end{align}

Under time-reversal symmetry ($T$), the pseudospin, momentum, and chirality transform as
\begin{align}
\vec{s} &\to -\vec{s}\nonumber\\
\vec{p} &\to -\vec{p}\\
\chi &\to \chi\nonumber
\end{align}

In a material that has both inversion and time-reversal symmetries, the chirality flips under $P$ and remains invariant under $T$. Crystal momentum $k_i$ flips sign under both $P$ and $T$. Therefore, under $PT$,
\begin{align}
k_i \to k_i\nonumber \\
\chi \to -\chi 
\end{align}
In such a material, the left- and right-handed fermions coincide in the Brillouin zone; thus, it has Dirac cones, not Weyl cones. Therfore, all Weyl materials lack time-reversal, inversion or both.

Symmetries that reverse spatial orientation transform left-handed fermions into right-handed fermion and vice versa.
Therefore, in materials with these symmetries, each left-handed cone has a right-handed partner cone at the same energy and with the same velocities.
Consequently, effects such as the quantized circular photogalvanic effect, which requires an asymmetry between left- and right-handed cones, are possible only in materials that have a chiral crystal lattice without orientation-reversing symmetries.

\begin{figure}[htp]
     \centering
         \includegraphics[width=\linewidth]{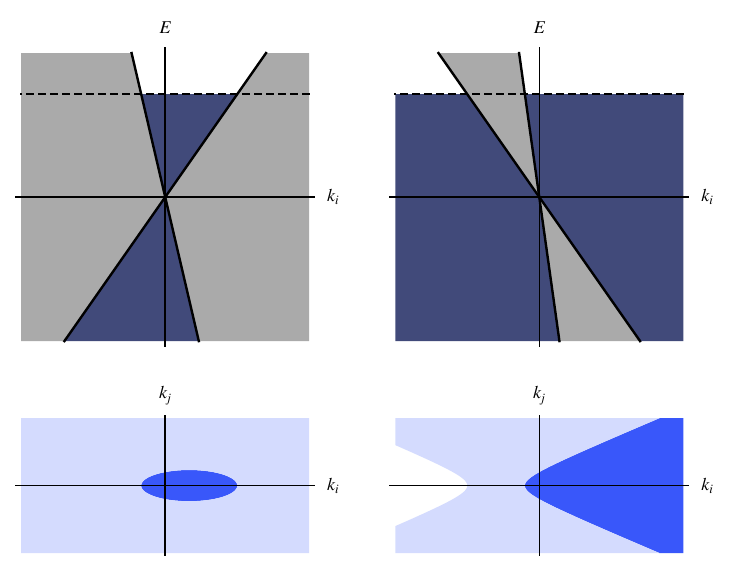}
        \caption{Dispersion relation (top) and Fermi surface (bottom) of type I (left) and type II (right) Weyl cones with tilt 0.5 and 1.5 respectively. In the upper figures, the dashed line shows the Fermi level, blue shows filled states, and white shows unfilled states; gray shows the region outside the cone. In the lower figures dark blue indicates both bands filled, light blue indicates one band filled, and white indicates both bands empty.}
        \label{types}
\end{figure}

The cones of Weyl fermions are generally tilted. Fermions with small tilts and elliptical Fermi surfaces are called type I Weyl fermions, while those with large tilts and hyperbolic Fermi surfaces are called type II Weyl fermions \cite{soluyanov2015}. The dispersion relations of type I and type II Weyl cones are sketched in Figure~\ref{types}. Many Weyl semimetals have exclusively type I fermions, such as TaAs~\cite{lv2015PRX,xu2015,yang2015,lv2015Nat}. Some materials have only type II fermions, such as $\mathrm{WTe_2}$~\cite{li2017} . The Weyl semimetal $\mathrm{OsC_2}$ is unusual in that it has both kinds of Weyl cones\textemdash24 of type I and 12 of type II \cite{zhang2018}. 


The linearized Hamiltonian for a general Weyl cone is
\begin{equation}
H = v_\text{t}^i q_i + v_a^i \sigma_a q_i + E,
\end{equation}
where $v_\text{t}^i$ is the ``tilt" velocity and $v^i_a$ represents the untilted part of the Hamiltonian. $E$ is the energy of the Weyl point. The matrices $\sigma_a$ act on spin or pseudospin. The chirality is
\begin{equation}\label{chi}
    \chi = \mathrm{sgn(det}(v^i_a)),
\end{equation} 
which is positive for a right-handed cone and negative for a left-handed cone. We define the product of velocities 
\begin{equation}\label{v1v2v3}
    v_1v_2v_3 = |\mathrm{det}(v^i_a)|,
\end{equation} 
the velocity tensor $(v_\text{W}^2)^{ij} = v_a^i v_a^j$ and its inverse $(v_\text{W}^{-2})_{ij}$. We also define a dimensionless measure of the tilt
\begin{equation}\label{tilt}
    \text{tilt parameter} = \sqrt{(v_\text{W}^{-2})_{ij}v_\text{t}^i v_\text{t}^j}.
\end{equation}The cone is type I if the tilt parameter is less than 1 and type II if it is greater than 1 \cite{soluyanov2015}.

 
\section{True and False chirality}

L. D. Barron introduced the idea of true and false chirality of a system \cite{barron1986}. 
A Weyl material is said to have false chirality if it possesses a symmetry $MT$, where $M$ is some symmetry that reverses spatial orientation, but does not have the symmetry $M$ itself.
Such a material transforms to its mirror image under time reversal. Systems with true chirality retain their chirality even under time reversal. Examples of systems with true chirality include glucose and DNA molecules and the electroweak part of the Standard Model. Systems with false chirality include a cone rotating about its axis and a crystal with inversion symmetry subjected to parallel uniform electric and magnetic fields.

Asymmetric Weyl materials, such as RhSi and CoSi, have crystal structures with true chirality. Materials with false chirality cannot be asymmetric Weyl materials, because the chirality of each cone is invariant under $T$ but flips under $M$ (since, by definition, $M$ reverses orientation), so left- and right-handed cones would be related by $MT$.

In symmetric Weyl materials, 
it is possible to break mirror symmetries by applying external perturbations. 
Such symmetry breaking may produce either true or false chirality, as we will demonstrate.
As a first example, consider the transition metal monopnitcide class of materials, which includes TaAs, TaP, NbAs, and NbP~\cite{lv2015PRX,xu2015,yang2015,lv2015Nat,xu2016, modic2019, xu2015discovery, yuan2020}.
These compounds have tetragonal symmetry and are in the space group $I4_1md$ (No. 109) with fourfold rotation about the [001] axis, reflection symmetry about the $[100], [010], [110], [\bar{1} 10]$ planes, and time-reversal symmetry, but no inversion symmetry or reflection symmetry about the $[001]$ plane.
If we apply a magnetic field along the $[001]$ direction, $B_z$ flips sign under reflections and therefore breaks all mirror symmetries. However, this perturbation does not introduce true chirality as $B_z$ also flips sign under time-reversal; therefore the perturbed system remains invariant under symmetries $MT$, where $M$ is a mirror reflection symmetry of the unperturbed system. Thus, TaAs with a magnetic field along the $c$ axis has false chirality and would therefore still be a symmetric Weyl material. 

Only systems with true chirality will display effects that depend on a difference in energy or velocity between left- and right-handed fermions.
Such effects will correspond to the expectation value of an operator, $Q$, such that $Q$ is invariant under all symmetries that preserve spatial orientation, but reverses sign under all symmetries that reverse spatial orientation (i.e. $Q$ has the same transformation as spatial orientation under the crystal symmetries).
For a perturbation $\lambda$ to produce true chirality in the system, there must exist a function $Q(\lambda)$ with this property.

For systems that possess time-reversal symmetry, a quantity that is of odd order in magnetic field does not qualify as $Q(\lambda)$ because it is odd under time-reversal, which preserves chirality. For example, in the transition metal monopnictides mentioned above, $B_z$ is invariant under rotation, and flips under reflections, but since it also flips under time-reversal, which preserves the chiralities of fermions, operators that are odd in $B_z$ will yield a vanishing expectation value.

\begin{figure}[htp]
  \includegraphics[width=0.9\linewidth]{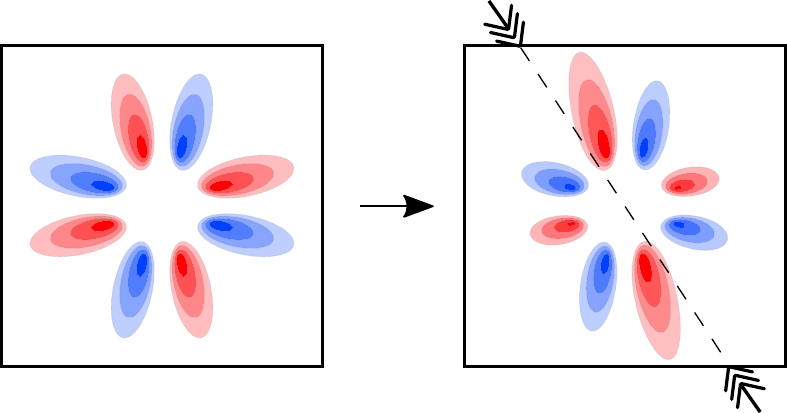}
  \caption{Slices of the Fermi surface for a Weyl material with fourfold rotation symmetry such as TaAs. The shading represents energy and the color represents chirality of the fermions. Each ellipse represents a Weyl cone. A perturbation along a low symmetry axis, such as a strain or a magnetic field, would affect left- and right-handed cones differently, resulting in a net chiral asymmetry.}
  \label{pert}
\end{figure}

In TaAs, the lowest order $Q$ as a function of magnetic field is $Q(\vec{B}) = B_x B_y (B_x^2 - B_y^2)$. Similarly, if we consider chirality induced by strain along a low symmetry axis, the lowest order $Q(S)$ is $S_{xy} (S_{xx} - S_{yy})$, where $S_{ij}$ is the strain tensor.
Therefore, in TaAs, we expect physical phenomena such as the helical magnetic effect to be of fourth order in $\vec{B}$ or second order in $S_{ij}$. 
The effects of a low symmetry perturbation are illustrated in Figure~\ref{pert}.
{As a second example, the rare earth carbides studied in \cite{ray2020} have the space group $Amm2$ (No. 38). In systems with this symmetry, the lowest order function of magnetization $\vec{\mu}$ that transforms in the same way as the chirality of fermions under all symmetries of the unperturbed system is $Q(\vec{\mu}) = \mu_x \mu_y$. }

In materials that possess inversion symmetry, external magnetic fields or uniform strain cannot induce chirality, because both magnetic field and strain are even under parity, and any functions of these quantities will also be even under parity. Since most Dirac materials such as $\mathrm{ZrTe_5}$ and $\mathrm{Na_3 Bi}$ have inversion symmetry, we cannot transform them into asymmetric Weyl materials through uniform strain or uniform external magnetic field. However, it is possible to create symmetric Weyl cones and other topological phases in materials such as $\mathrm{ZrTe_5}$ with Zeeman splitting \cite{sun2020topological,choi2020}.

We now specialize to the case of breaking the chiral symmetry in a material with no magnetic ordering by a uniform magnetic field. Since this magnetic field is translation invariant, the only symmetries that can change the magnetic field are the symmetries of the point group, and their products with time reversal. Therefore, the expressions $Q(\vec{B})$ depend only on the point group, rather than the whole space group.
Out of the 32 point groups, 11 are chiral and already asymmetric.
Another 11 are centrosymmetric; as discussed in the previous paragraph, magnetic field cannot induce chirality in these groups.
In the other 10 groups, we can use a magnetic field to break the chiral symmetry.
We list the function $Q(\vec{B})$ in Table~\ref{QB} for these 10 groups. 
The symmetry analysis in Table~\ref{QB} holds only for perturbations that can be described by uniform time-odd pseudovectors, such as a uniform magnetic field and ferromagnetism.

But a magnetic field is not the only perturbation that can create asymmetry in Weyl materials. For example, Weyl points can be created or manipulated by strain \cite{ruan2016, cortijo2016}, magnetic ordering \cite{ray2020,shekhar2018,ghimire2019}, incident light \cite{hubener2017}, ferroelectricity \cite{he2018,sharma2019}, or a superconducting condensate \cite{obrien2017}. The symmetry analysis in Table~\ref{QB} is not applicable to these perturbations. For example, an electric polarization along a low symmetry direction, which is described by a time-even vector, can break chiral symmetry even in materials with the $O_h$ point group. The relevant parameter that breaks chiral symmetry is $Q(\vec{P}) = P_x P_y P_z (P_x^2 - P_y^2)(P_y^2 - P_z^2)(P_z^2 - P_x^2)$.

\begin{table}[htp]
\centering
\begin{tabular}{c c}
\toprule
Group & $Q(\vec{B})$ \\
\midrule
$m(C_s)$ & $B_x B_z$ \textbf{OR} $B_y B_z$ \\
$mm2(C_{2v})$ & $B_x B_y$\\
$\bar{4}(S_4)$ & $B_x B_y$ \textbf{OR} $B_x^2 - B_y^2$\\
$4mm(C_{4v})$ & $B_x B_y (B_x^2 - B_y^2)$\\
$\bar{4}2m(D_{2d})$ & $B_x^2 - B_y^2$\\
$3m(C_{3v})$ & $(B_1 - B_2)(B_2 - B_3)(B_3 - B_1)(B_1 + B_2 + B_3)$\\
$\bar{6}(C_{3h})$ & $B_z B_a B_b B_c$ \textbf{OR} $B_z (B_a - B_b)(B_b - B_c)(B_c - B_a)$\\
$6mm(C_{6v})$ & $B_a B_b B_c (B_a - B_b)(B_b - B_c)(B_c - B_a)$\\
$\bar{6}m2(D_{3h})$ & $B_z B_a B_b B_c$\\
$\bar{4}3m(T_d)$ & $(B_x^2 - B_y^2)(B_y^2 - B_z^2)(B_z^2 - B_x^2)$\\
\bottomrule
\end{tabular}
\caption{\label{QB} $Q(\vec{B})$ for the 10 point groups where $\vec{B}$ can break chirality. 
In the first five rows and the last row, $x, y, z$ indicate the orthorhombic crystal axes; 
in the sixth row, $1,2,3$ indicate the rhombohedral crystal axes;
in the remaining rows, $z, a, b, c$ indicate hexagonal lattice vectors, with $\hat{a} + \hat{b} + \hat{c} = \vec{0}$. 
When multiple functions $Q$ are listed, it is enough for one of these quantities to be non-zero to break chiral symmetry.}
\end{table}

In this work, we focus on a class of non-centrosymmetric materials that have the space group $F\bar{4}3m$ (No. 216) and the point group $\bar{4}3m (T_d)$. This class includes the half-Heusler compound GdPtBi, which is antiferromagnetic at low temperatures \cite{hirschberger2016} and the zincblende compounds HgTe and InSb. In a state without magnetic ordering, these materials have several axes of rotation and planes of reflection, as well as time-reversal symmetry, but they lack inversion symmetry. Their electronic structure is characterized by a fourfold degeneracy at the $\Gamma$ point and no Weyl cones. However, if we apply an external magnetic field~\cite{cano2017} or  strain~\cite{ruan2016}, or if there is magnetic ordering in the case of GdPtBi,~\cite{hirschberger2016,shekhar2018}, the degeneracy splits, and Weyl points appear. When the degeneracy is broken by a magnetic field along high symmetry axes, such as [111] and [100], all the emergent Weyl fermions come in pairs of opposite chirality related by symmetry \cite{cano2017}.
In the next section, we will show that this is not the case when the magnetic field is along a low-symmetry axis.


\section{Model}

We consider the Hamiltonian given in Ref.~\cite{cano2017}:
\begin{align}\label{Ham}
H = A k^2 I_4 + &C\left[(k_x^2 - k_y^2)\Gamma_1 + \frac{1}{\sqrt{3}}(2k_z^2 - k_x^2 - k_x^2)\Gamma_2\right]\nonumber\\  + &F(k_x k_y \Gamma_3 + k_x k_z \Gamma_4 + k_y k_z \Gamma_5)\nonumber\\ + &D(k_x U_x + k_y U_y + k_z U_z)\nonumber\\
+& \mu_B g (B_x J_x + B_y J_y + J_z B_z),
\end{align}
where $J_{x,y,z}$ are the spin-$\frac{3}{2}$ matrices, which can be expressed as
\begin{align}
J_x &= 
\begin{psmallmatrix}
 0 & \frac{\sqrt{3}}{2} & 0 & 0 \\
 \frac{\sqrt{3}}{2} & 0 & 1 & 0 \\
 0 & 1 & 0 & \frac{\sqrt{3}}{2} \\
 0 & 0 & \frac{\sqrt{3}}{2} & 0 \\
\end{psmallmatrix},  \\
J_y &= 
\begin{psmallmatrix}
 0 & -\frac{i \sqrt{3}}{2} & 0 & 0 \\
 \frac{i \sqrt{3}}{2} & 0 & -i & 0 \\
 0 & i & 0 & -\frac{i \sqrt{3}}{2} \\
 0 & 0 & \frac{i \sqrt{3}}{2} & 0 \\
\end{psmallmatrix},  \\
J_z &= 
\begin{psmallmatrix}
 \frac{3}{2} & 0 & 0 & 0 \\
 0 & \frac{1}{2} & 0 & 0 \\
 0 & 0 & -\frac{1}{2} & 0 \\
 0 & 0 & 0 & -\frac{3}{2} \\
\end{psmallmatrix}
\end{align}

The matrices $\Gamma_\mu$ are defined as
\begin{align}
\Gamma_1 &= \frac{1}{\sqrt{3}}(J_x^2 - J_y^2)\nonumber\\
\Gamma_2 &= \frac{1}{3}(2J_z^2 - J_x^2 - J_y^2)\nonumber\\
\Gamma_3 &= \frac{1}{\sqrt{3}}\{J_x,J_y\}\\
\Gamma_4 &= \frac{1}{\sqrt{3}}\{J_x,J_z\}\nonumber\\
\Gamma_5 &= \frac{1}{\sqrt{3}}\{J_y,J_z\}\nonumber
\end{align}
These matrices satisfy the anticommutation relations $\{\Gamma_\mu,\Gamma_\nu\} = 2\delta_{\mu\nu}$. The matrices $U_i$ are defined as
\begin{align}
U_x &= \frac{1}{2i}(\sqrt{3}[\Gamma_1,\Gamma_5] - [\Gamma_2,\Gamma_5])\nonumber\\
U_y &= \frac{-1}{2i}(\sqrt{3}[\Gamma_1,\Gamma_4] + [\Gamma_2,\Gamma_4])\\
U_z &= \frac{1}{i}[\Gamma_2,\Gamma_3]\nonumber
\end{align}
%
The unperturbed Hamiltonian (at zero magnetic field) has time-reversal and $T_d$ symmetry. The coefficient $D$ represents an inversion breaking term. 

We seek an expression $Q(\vec{B})$ that indicates true chirality. As discussed in the previous section, $Q(\vec{B})$ must be even(odd) under symmetries that maintain(reverse) the chirality of a Weyl cone.
The expression $B_x B_y B_z$ represents false chirality because it is odd under time-reversal and even under $S_4T$, where $S_4$ is the rotoinversion symmetry about the $z$ axis that maps $(x,y,z) \mapsto (y,-x,-z)$.
The lowest order polynomial of $B$ that breaks true chiral symmetry is $Q(\vec{B}) = (B_x^2 - B_y^2)(B_y^2 - B_z^2)(B_z^2 - B_x^2)$, as listed in Table~\ref{QB}. It is non-zero only when the magnetic field is away from high symmetry axes such as [100], [110], and [111]. 
Because the magnetic field itself lifts the degeneracy at the $\Gamma$ point and generates Weyl points, in what follows, we will use the dimensionless quantity
\begin{equation}\label{qprime}
    Q'(\hat{B}) = (B_x^2 - B_y^2)(B_y^2 - B_z^2)(B_z^2 - B_x^2)/B^6
\end{equation}
to measure how much chiral symmetry is broken, i.e. how much the left and right handed Weyl points might differ by each other.  $Q'(\hat{B})$ takes a maximum value of $ 0.0962$ when $\hat{B} = (0, 0.460, 0.888)$ or points in a symmetry-related direction.


The Hamiltonian defined by Eq.~\eqref{Ham} is a lowest order expansion in $k$ that is only valid for a finite range of crystal momentum and energy; at higher energies, mixing with other bands becomes important. 
Therefore, when we search for Weyl points in this Hamiltonian with an applied magnetic field, we choose a cut-off in crystal momentum and energy and only focus on Weyl points that are within this cutoff.
We ensure that the cutoff always contains the same number of left- and right-handed Weyl points.

We will now focus on the zincblende material InSb.
It has six bands closest to the Fermi level, consisting of a set of four valence bands that meet at $\Gamma$ very close to $E=0$ and a set of two conduction bands separated by a bandgap of 235 meV. 
The set of four is modeled by the Hamiltonian in Eq.~\eqref{Ham} with parameters as given in Table~\ref{parameters} \cite{cano2017,qu2016,nilsson2009,vurgaftman2001}.
\begin{table}[htp]
\centering
\begin{tabular}{lr}
\toprule
   $A$ & $\SI{8.85}{eV \angstrom^2}$\\
   $C$ & $\SI{14.2}{eV \angstrom^2}$\\
   $D$ & $\SI{0.01}{eV \angstrom}$\\
   $F$ & $\SI{-22}{eV \angstrom^2}$\\
   $g$-factor & $51$\\
\bottomrule
\end{tabular}
\caption{\label{parameters} Hamiltonian parameters in Eq.~\eqref{Ham} corresponding to InSb.}
\end{table}

\section{Weyl fermions in indium antimonide in a magnetic field}
When a magnetic field is applied along a general low-symmetry direction, the induced Weyl points are not related to each other by any symmetry. There is no analytical expression for their positions and each one must be located numerically by diagonalizing Eq.~\eqref{Ham}. 
We will search for Weyl points within the cutoffs $k < \SI{0.032}{\angstrom^{-1}}$ and $E < \SI{50}{meV}$. We focus on magnetic fields $\ge \SI{0.5}{T}$, because for very small magnetic fields, the separation bewteen the Weyl points in momentum space and difference between the energies of the Weyl points and other bands are too small.  For all magnitudes and directions of magnetic field considered in our work, there is an equal number of left- and right-handed Weyl cones within these cutoffs. 

In addition to $Q'(\hat{B})$, we characterize the breaking of chiral symmetry by a dimensionless parameter 
\begin{equation}\label{deltav}
\delta_v = \sum\chi v_1v_2v_3/\sum v_1v_2v_3, 
\end{equation}
where the product of velocities $v_1 v_2 v_3$ is defined in Eq.~\eqref{v1v2v3} and the sum is over all the cones. We also characterize by how much chiral symmetry is broken by a dimensionful parameter 
\begin{equation}\label{deltaE}
\delta_E = \sum\chi E/n    
\end{equation}
where $n$ is the total number of right-handed Weyl cones (equal to the total number of left-handed Weyl cones). Physically, $\delta_v$ represents the average difference in velocities of the left- and right-handed Weyl cones normalized by the average velocity, while $\delta_E$ represents the average difference in energy between left- and right-handed Weyl cones.

Because $\delta_v$ is a dimensionless parameter, we expect it to vary strongly with $Q'(\hat{B})$ (which depends only on the direction of $\vec{B}$) and weakly, if at all, with the magnitude of $
\vec{B}$. However $\delta_E$, which is a dimensionful parameter, is expected to vary strongly with both $Q'(\hat{B})$ and the magnitude of $\vec{B}$.

While $Q'(\hat{B})$ serves as a quick check to determine which low-symmetry directions are likely to break chiral symmetry the most, $\delta_v$ and $\delta_E$ are directly related to known physical observables  \cite{dejuan2017,yuta2018,meyer2018}. Note that $Q'(\hat{B})\neq 0$ is a necessary condition for the left- and right- handed cones to have different energies and velocities, and thus for $\delta_v$, $\delta_E$, and relevant physical observables to be non-zero.

In Table~\ref{detailedtable}, we have listed the energy, momentum, chirality, velocity and tilt of all ten Weyl cones that appear for a magnetic field of 0.75 T along the low symmetry direction $[147]$. The parameter $Q'(\hat{B})$ for this direction is 0.0826, close to the maximum possible value of 0.0962. 
For this field, we observe three type I Weyl cones and seven type II Weyl cones: unexpectedly, the type I (or type II) Weyl cones do not come in pairs of opposite chirality, which is only possible in asymmetric Weyl materials.

In Table~\ref{magtable}, we have listed the number of type I and type II cones of each chirality, and the parameters $\delta_v$ and $\delta_E$, for different magnitudes of magnetic field along the $[147]$ direction.
The values show that $\delta_v$, the dimensionless normalized average velocity difference between left- and right-handed Weyl cones, increases slightly with increasing magnitude of the magnetic field, even while $Q'(\hat{B})$ remains constant. The average energy difference, $\delta_E$, increases strongly with increasing field.
Thus, we expect physical observables that depend on a difference in energy or velocity between Weyl cones will increase as the magnetic field is increased along this direction.

In addition, Table~\ref{magtable} also shows that while the number of left and right handed cones of each chirality remains the same as we change the magnitude of the magnetic field, the number of type I (and type II) right handed cones changes. Type I and type II cones have very different Fermi surfaces as shown in Figure \ref{types}. When the tilt parameter is close to $1$, even a small change in the parameters of the Hamiltonian results in a drastic change in the Fermi surface.

The same quantities are recorded in Table~\ref{dirtable} for different directions of magnetic field and fixed magnitude 0.75~T.
Here the three quantities that characterize breaking of chirality symmetry, $Q'(\hat{B})$, $\delta_v$ and $\delta_E$ can be compared: $\delta_v$ and $\delta_E$ show similar trends as $Q'(\hat{B})$, but are not functions of $Q'(\hat{B})$.
Because of the complexities of the band structure and topology, they do not even necessarily vary monotonically with $Q(\vec{B})$ or $Q'(\hat{B})$: they can change sign or even be zero for some low-symmetry directions of $\vec{B}$.
Both Tables~\ref{magtable} and \ref{dirtable} show that it is not unusual to find different numbers of left- and right-handed Weyl cones of the same type in this model.


%
%

\begin{figure}[htp]
     \centering
     \begin{subfigure}[b]{0.3\linewidth}
         \centering
         \includegraphics[width=\textwidth]{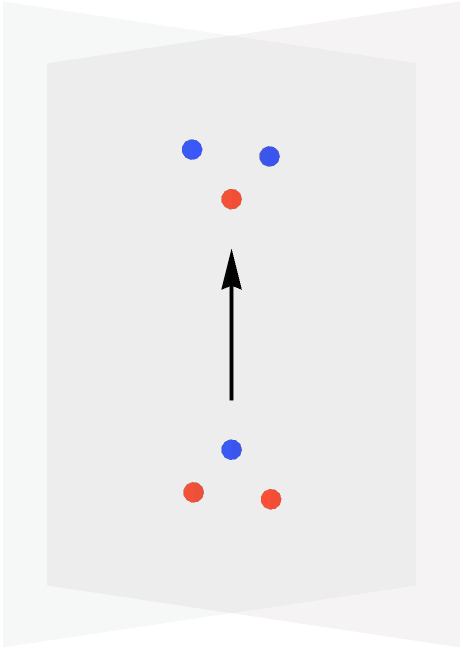}
         \caption{$[001]$}
         \label{fig:cones1}
     \end{subfigure}\hspace{0.03\linewidth}
     \begin{subfigure}[b]{0.3\linewidth}
         \centering
         \includegraphics[width=\textwidth]{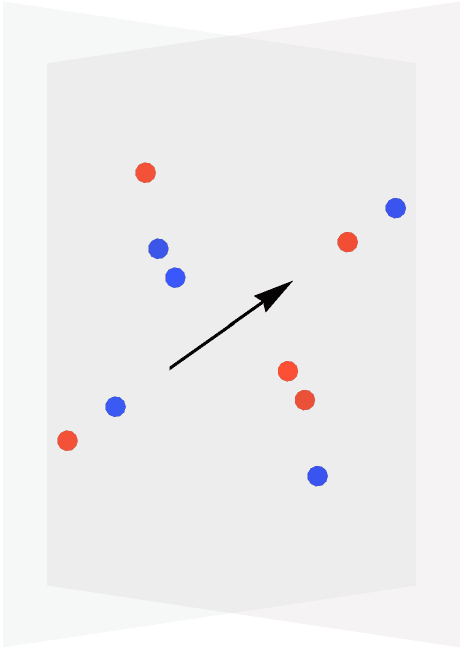}
         \caption{$[111]$}
         \label{fig:cones2}
     \end{subfigure}\hspace{0.03\linewidth}
     \begin{subfigure}[b]{0.3\linewidth}
         \centering
         \includegraphics[width=\textwidth]{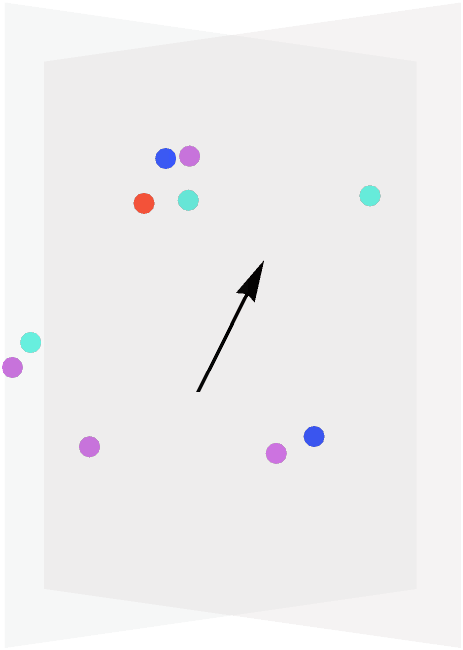}
         \caption{$[147]$}
         \label{fig:cones3}
     \end{subfigure}
        \caption{Distribution of Weyl points in momentum space for a magnetic field of 0.75 T along high symmetry directions (a) $[001]$ and (b) $[111]$ and the low symmetry direction (c) $[147]$. The arrow denotes the direction of magnetic field. The color represents the chirality and type of the cones: red indicates left-handed type I, magenta indicates left-handed type II, blue indicates right-handed type I, and cyan indicates right-handed type II. For the low symmetry direction $[147]$, there is no symmetry relating the cones.}
        \label{fig:cones}
\end{figure}

As an example, the positions of cones of different types and chiralities for a magnetic field of 0.75 T along the directions $[001]$, $[111]$, and $[147]$ are shown in Figure~\ref{fig:cones}.



\begin{table}[htp]
\centering
\begin{tabular}{rrrcrcrrc}
\toprule
\multicolumn{1}{c}{$k_x$} & \multicolumn{1}{c}{$k_y$} & \multicolumn{1}{c}{$k_z$} & \multicolumn{1}{c}{$\chi$} & \multicolumn{1}{c}{Energy} & \multicolumn{1}{c}{Bands} & \multicolumn{1}{c}{$v_1 v_2 v_3$} & \multicolumn{1}{c}{Tilt} & \multicolumn{1}{c}{Type} \\
\multicolumn{3}{c}{\multirow{2}{*}{(\SI{e8}{m^{-1}})}} &  \hspace{0.5cm} & \multicolumn{1}{c}{\multirow{2}{*}{(meV)}} & & \multicolumn{1}{c}{$(\si{eV^3}$} & \multicolumn{1}{c}{} & \multicolumn{1}{c}{} \\
\multicolumn{6}{c}{} & \multicolumn{1}{c}{$\si{pm^3})$} & \multicolumn{2}{c}{} \\
\midrule
-0.229 & -0.569 & 0.782  & L & -1.36 & 1-2 & 5830  & 0.706 & I \\
-0.755 & -0.572 & -0.802 & L & 1.82  & 2-3 & 13200 & 1.19  & II \\
0.812  & -0.432 & -0.782 & L & 1.32  & 2-3 & 9350  & 1.20  & II \\
-0.083 & -0.302 & 1.093  & L & 0.20  & 2-3 & 1780  & 1.87  & II \\
-2.114 & -0.120 & -0.304 & L & 11.72 & 3-4 & 12.6  & 333   & II \\
-0.016 & -0.579 & 1.058  & R & 0.10  & 2-3 & 2950  & 0.585 & I \\
0.310  & 0.462  & -0.733 & R & -1.59 & 1-2 & 4390  & 0.710 & I \\
0.735  & 0.533  & 0.837  & R & 1.81  & 2-3 & 11400 & 1.32  & II \\
-0.762 & 0.344  & 0.865  & R & 1.30  & 2-3 & 6430  & 1.57  & II \\
-1.648 & -0.335 & -0.119 & R & 7.45  & 3-4 & 44.7  & 40.1  & II\\     
\bottomrule
\end{tabular}
\caption{\label{detailedtable} Weyl cones for a magnetic field of $0.75\mathrm{T}$ along the low symmetry direction $[147]$. 
The first three columns specify the crystal momentum of the Weyl point; the next columns indicate its chirality; its energy; the two bands that comprise it, where band 1 has the lowest energy; its product of velocities; its tilt; and whether it is type 1 or type II.
The chirality $\chi$, product of velocities $v_1 v_2 v_3$ and tilt are defined in Eqs.~\eqref{chi},~\eqref{v1v2v3}, and~\eqref{tilt}, respectively.
}
\end{table}

\begin{table}[htp]
\centering
\begin{tabular}{rccccrr}\toprule
\multicolumn{1}{c}{$B$} & \multicolumn{2}{c}{Type I} & \multicolumn{2}{c}{Type II} & \multicolumn{1}{c}{$\delta_v$} & \multicolumn{1}{c}{$\delta_E$} \\
\multicolumn{1}{c}{(T)} & \multicolumn{1}{c}{Left} & \multicolumn{1}{c}{Right} & \multicolumn{1}{c}{Left} & \multicolumn{1}{c}{Right} & \multicolumn{1}{c}{} & \multicolumn{1}{c}{(meV)} \\ \midrule
0.500                   & 1                          & 2                           & 4                           & 3                            & -0.0820                          & -0.39                                \\
0.625                 & 1                          & 2                           & 4                           & 3                            & -0.0863                          & -0.64                                \\
0.750                  & 1                          & 2                           & 4                           & 3                            & -0.0907                          & -0.92                                \\
0.875                 & 1                          & 2                           & 4                           & 3                            & -0.0953                          & -1.25                                \\
1.000                     & 1                          & 1                           & 4                           & 4                            & -0.1012                          & -1.61     \\                          \bottomrule
\end{tabular}
\caption{\label{magtable} Number of left- and right-handed cones of each type and parameters that characterize by how much chiral symmetry is broken for variable magnetic field $B$ along the low symmetry direction $[147]$. The parameters $\delta_v$ and $\delta_E$ are defined by Eq.~\eqref{deltav} and Eq.~\eqref{deltaE} respectively.}
\end{table}

\begin{table}[htp]
\centering
\begin{tabular}{lrccccrr}\toprule
\multicolumn{1}{c}{Dir} & \multicolumn{1}{c}{$Q'(\hat{B})$} & \multicolumn{2}{c}{Type I} & \multicolumn{2}{c}{Type II} & \multicolumn{1}{c}{$\delta_v$} & \multicolumn{1}{c}{$\delta_E$} \\
& \multicolumn{1}{c}{} & \multicolumn{1}{c}{Left} & \multicolumn{1}{c}{Right} & \multicolumn{1}{c}{Left} & \multicolumn{1}{c}{Right} & \multicolumn{1}{c}{} & \multicolumn{1}{c}{(meV)} \\ \midrule
{[}111{]} & 0                     & 5                          & 5                           & 0                           & 0                            & 0                               & 0                                    \\
{[}345{]} & 0.00806               & 2                          & 3                           & 2                           & 1                            & 0.00042                         & 0.0034                               \\
{[}123{]} & 0.0437                & 1                          & 2                           & 4                           & 3                            & -0.0175                          & -0.26                                \\
{[}147{]} & 0.0826                & 1                          & 2                           & 4                           & 3                            & -0.0907                          & -0.92                                \\
{[}001{]} & 0                     & 3                          & 3                           & 0                           & 0                            & 0                               & 0    \\
\bottomrule
\end{tabular}
\caption{\label{dirtable} Number of left- and right-handed cones of each type and parameters characterize by how much chiral symmetry is broken for magnetic field of magnitude 0.75 T along different directions (Dir). The parameters $Q'$, $\delta_v$ and $\delta_E$ are defined by Eq.~\eqref{qprime}, Eq.~\eqref{deltav} and Eq.~\eqref{deltaE} respectively. }
\end{table}

\section{Outlook}

Asymmetric Weyl materials are sought-after to observe effects that require breaking of chiral symmetry, such as the quantized circular photogalvanic effect~\cite{dejuan2017}, the helical magnetic effect~\cite{yuta2018}, and the chiral magnetic effect without an external source of chirality~\cite{meyer2018}. Yet few of these materials have been shown to exist naturally. Therefore, in this work, we proposed inducing an asymmetry between the left- and right-handed Weyl cones in otherwise symmetric materials by an applied field, such as strain or magnetic field, along a low-symmetry direction. We have also provided a prescription for distinguishing true from false chirality, namely, the existence of an operator $Q$ that is even(odd) under all chirality-preserving(chirality-flipping) symmetries of the crystal.

We then studied how to induce true chirality in materials with $T_d$ symmetry. We introduced a parameter $Q'(\hat{B})$ which determines whether chiral symmetry is broken. We applied this analysis to the specific case of InSb. There we showed, by exact diagonalization of a low-energy model, that for a magnetic field along low symmetry directions, the energies, velocities, and tilts are different for left- and right-handed Weyl cones. We also introduced the parameters $\delta_v$ and $\delta_E$ to quantify the asymmetry between the left- and right-handed cones. The differences in energy will lead to the quantized circular photogalvanic effect, while differences in tilt and velocities will lead to the chiral magnetic effect without external source of chirality and the helical magnetic effect. 


For several directions and magnitudes of the magnetic field, the number of left- and right-handed type I cones (and number of left- and right-handed type II cones) are different. Of course, the total number of left- and right-handed cones remains equal, as required by topology. Since type I cones have a compact Fermi surface, and type II cones have a hyperbolic Fermi surface, this asymmetry between right- and left-handed Weyl cones of the same type will result in a highly nontrivial topology of Berry curvature.

Unlike intrinsically asymmetric Weyl materials, in the materials discussed in this manuscript, effects that depend on breaking chiral symmetry can be turned on and off, and flipped in sign, which may be desirable for measuring certain effects. Our results can be generalized to other space groups and different types of symmetry-breaking perturbations.

\begin{acknowledgments}
We thank Dmitri Kharzeev, Gao Lanlan, and Mengkun Liu for useful and stimulating discussions. This work was supported in part by the U. S. Department of Energy under Awards DE-SC-0017662 (S. K.) and DE-FG02-88ER40388 (E. J. P.) and by the National Science Foundation under award DMR-1942447 (J. C.).
J. C. acknowledges the support of the Flatiron Institute, a division of the Simons Foundation.
J. C. and S. K. also acknowledge the support of an OVPR Seed Grant from Stony Brook University.
\end{acknowledgments}

\bibliographystyle{apsrev4-1}
\bibliography{bibliography}

\begin{thebibliography}{62}%
\makeatletter
\providecommand \@ifxundefined [1]{%
 \@ifx{#1\undefined}
}%
\providecommand \@ifnum [1]{%
 \ifnum #1\expandafter \@firstoftwo
 \else \expandafter \@secondoftwo
 \fi
}%
\providecommand \@ifx [1]{%
 \ifx #1\expandafter \@firstoftwo
 \else \expandafter \@secondoftwo
 \fi
}%
\providecommand \natexlab [1]{#1}%
\providecommand \enquote  [1]{``#1''}%
\providecommand \bibnamefont  [1]{#1}%
\providecommand \bibfnamefont [1]{#1}%
\providecommand \citenamefont [1]{#1}%
\providecommand \href@noop [0]{\@secondoftwo}%
\providecommand \href [0]{\begingroup \@sanitize@url \@href}%
\providecommand \@href[1]{\@@startlink{#1}\@@href}%
\providecommand \@@href[1]{\endgroup#1\@@endlink}%
\providecommand \@sanitize@url [0]{\catcode `\\12\catcode `\$12\catcode
  `\&12\catcode `\#12\catcode `\^12\catcode `\_12\catcode `\%12\relax}%
\providecommand \@@startlink[1]{}%
\providecommand \@@endlink[0]{}%
\providecommand \url  [0]{\begingroup\@sanitize@url \@url }%
\providecommand \@url [1]{\endgroup\@href {#1}{\urlprefix }}%
\providecommand \urlprefix  [0]{URL }%
\providecommand \Eprint [0]{\href }%
\providecommand \doibase [0]{http://dx.doi.org/}%
\providecommand \selectlanguage [0]{\@gobble}%
\providecommand \bibinfo  [0]{\@secondoftwo}%
\providecommand \bibfield  [0]{\@secondoftwo}%
\providecommand \translation [1]{[#1]}%
\providecommand \BibitemOpen [0]{}%
\providecommand \bibitemStop [0]{}%
\providecommand \bibitemNoStop [0]{.\EOS\space}%
\providecommand \EOS [0]{\spacefactor3000\relax}%
\providecommand \BibitemShut  [1]{\csname bibitem#1\endcsname}%
\let\auto@bib@innerbib\@empty
\bibitem [{\citenamefont {{Young}}\ \emph {et~al.}(2012)\citenamefont
  {{Young}}, \citenamefont {{Zaheer}}, \citenamefont {{Teo}}, \citenamefont
  {{Kane}}, \citenamefont {{Mele}},\ and\ \citenamefont {{Rappe}}}]{Young12}%
  \BibitemOpen
  \bibfield  {author} {\bibinfo {author} {\bibfnamefont {S.~M.}\ \bibnamefont
  {{Young}}}, \bibinfo {author} {\bibfnamefont {S.}~\bibnamefont {{Zaheer}}},
  \bibinfo {author} {\bibfnamefont {J.~C.~Y.}\ \bibnamefont {{Teo}}}, \bibinfo
  {author} {\bibfnamefont {C.~L.}\ \bibnamefont {{Kane}}}, \bibinfo {author}
  {\bibfnamefont {E.~J.}\ \bibnamefont {{Mele}}}, \ and\ \bibinfo {author}
  {\bibfnamefont {A.~M.}\ \bibnamefont {{Rappe}}},\ }\href {\doibase
  10.1103/PhysRevLett.108.140405} {\bibfield  {journal} {\bibinfo  {journal}
  {\prl}\ }\textbf {\bibinfo {volume} {108}},\ \bibinfo {eid} {140405}
  (\bibinfo {year} {2012})},\ \Eprint {http://arxiv.org/abs/1111.6483}
  {arXiv:1111.6483 [cond-mat.mes-hall]} \BibitemShut {NoStop}%
\bibitem [{\citenamefont {{Wang}}\ \emph {et~al.}(2012)\citenamefont {{Wang}},
  \citenamefont {{Sun}}, \citenamefont {{Chen}}, \citenamefont {{Franchini}},
  \citenamefont {{Xu}}, \citenamefont {{Weng}}, \citenamefont {{Dai}},\ and\
  \citenamefont {{Fang}}}]{Wang12}%
  \BibitemOpen
  \bibfield  {author} {\bibinfo {author} {\bibfnamefont {Z.}~\bibnamefont
  {{Wang}}}, \bibinfo {author} {\bibfnamefont {Y.}~\bibnamefont {{Sun}}},
  \bibinfo {author} {\bibfnamefont {X.-Q.}\ \bibnamefont {{Chen}}}, \bibinfo
  {author} {\bibfnamefont {C.}~\bibnamefont {{Franchini}}}, \bibinfo {author}
  {\bibfnamefont {G.}~\bibnamefont {{Xu}}}, \bibinfo {author} {\bibfnamefont
  {H.}~\bibnamefont {{Weng}}}, \bibinfo {author} {\bibfnamefont
  {X.}~\bibnamefont {{Dai}}}, \ and\ \bibinfo {author} {\bibfnamefont
  {Z.}~\bibnamefont {{Fang}}},\ }\href {\doibase 10.1103/PhysRevB.85.195320}
  {\bibfield  {journal} {\bibinfo  {journal} {\prb}\ }\textbf {\bibinfo
  {volume} {85}},\ \bibinfo {eid} {195320} (\bibinfo {year} {2012})},\ \Eprint
  {http://arxiv.org/abs/1202.5636} {arXiv:1202.5636 [cond-mat.mtrl-sci]}
  \BibitemShut {NoStop}%
\bibitem [{\citenamefont {{Liu}}\ \emph
  {et~al.}(2014{\natexlab{a}})\citenamefont {{Liu}}, \citenamefont {{Zhou}},
  \citenamefont {{Zhang}}, \citenamefont {{Wang}}, \citenamefont {{Weng}},
  \citenamefont {{Prabhakaran}}, \citenamefont {{Mo}}, \citenamefont {{Shen}},
  \citenamefont {{Fang}}, \citenamefont {{Dai}}, \citenamefont {{Hussain}},\
  and\ \citenamefont {{Chen}}}]{Liu14}%
  \BibitemOpen
  \bibfield  {author} {\bibinfo {author} {\bibfnamefont {Z.~K.}\ \bibnamefont
  {{Liu}}}, \bibinfo {author} {\bibfnamefont {B.}~\bibnamefont {{Zhou}}},
  \bibinfo {author} {\bibfnamefont {Y.}~\bibnamefont {{Zhang}}}, \bibinfo
  {author} {\bibfnamefont {Z.~J.}\ \bibnamefont {{Wang}}}, \bibinfo {author}
  {\bibfnamefont {H.~M.}\ \bibnamefont {{Weng}}}, \bibinfo {author}
  {\bibfnamefont {D.}~\bibnamefont {{Prabhakaran}}}, \bibinfo {author}
  {\bibfnamefont {S.~K.}\ \bibnamefont {{Mo}}}, \bibinfo {author}
  {\bibfnamefont {Z.~X.}\ \bibnamefont {{Shen}}}, \bibinfo {author}
  {\bibfnamefont {Z.}~\bibnamefont {{Fang}}}, \bibinfo {author} {\bibfnamefont
  {X.}~\bibnamefont {{Dai}}}, \bibinfo {author} {\bibfnamefont
  {Z.}~\bibnamefont {{Hussain}}}, \ and\ \bibinfo {author} {\bibfnamefont
  {Y.~L.}\ \bibnamefont {{Chen}}},\ }\href {\doibase 10.1126/science.1245085}
  {\bibfield  {journal} {\bibinfo  {journal} {Science}\ }\textbf {\bibinfo
  {volume} {343}},\ \bibinfo {pages} {864} (\bibinfo {year}
  {2014}{\natexlab{a}})},\ \Eprint {http://arxiv.org/abs/1310.0391}
  {arXiv:1310.0391 [cond-mat.mtrl-sci]} \BibitemShut {NoStop}%
\bibitem [{\citenamefont {{Liu}}\ \emph
  {et~al.}(2014{\natexlab{b}})\citenamefont {{Liu}}, \citenamefont {{Jiang}},
  \citenamefont {{Zhou}}, \citenamefont {{Wang}}, \citenamefont {{Zhang}},
  \citenamefont {{Weng}}, \citenamefont {{Prabhakaran}}, \citenamefont {{Mo}},
  \citenamefont {{Peng}}, \citenamefont {{Dudin}}, \citenamefont {{Kim}},
  \citenamefont {{Hoesch}}, \citenamefont {{Fang}}, \citenamefont {{Dai}},
  \citenamefont {{Shen}}, \citenamefont {{Feng}}, \citenamefont {{Hussain}},\
  and\ \citenamefont {{Chen}}}]{Liu14a}%
  \BibitemOpen
  \bibfield  {author} {\bibinfo {author} {\bibfnamefont {Z.~K.}\ \bibnamefont
  {{Liu}}}, \bibinfo {author} {\bibfnamefont {J.}~\bibnamefont {{Jiang}}},
  \bibinfo {author} {\bibfnamefont {B.}~\bibnamefont {{Zhou}}}, \bibinfo
  {author} {\bibfnamefont {Z.~J.}\ \bibnamefont {{Wang}}}, \bibinfo {author}
  {\bibfnamefont {Y.}~\bibnamefont {{Zhang}}}, \bibinfo {author} {\bibfnamefont
  {H.~M.}\ \bibnamefont {{Weng}}}, \bibinfo {author} {\bibfnamefont
  {D.}~\bibnamefont {{Prabhakaran}}}, \bibinfo {author} {\bibfnamefont {S.~K.}\
  \bibnamefont {{Mo}}}, \bibinfo {author} {\bibfnamefont {H.}~\bibnamefont
  {{Peng}}}, \bibinfo {author} {\bibfnamefont {P.}~\bibnamefont {{Dudin}}},
  \bibinfo {author} {\bibfnamefont {T.}~\bibnamefont {{Kim}}}, \bibinfo
  {author} {\bibfnamefont {M.}~\bibnamefont {{Hoesch}}}, \bibinfo {author}
  {\bibfnamefont {Z.}~\bibnamefont {{Fang}}}, \bibinfo {author} {\bibfnamefont
  {X.}~\bibnamefont {{Dai}}}, \bibinfo {author} {\bibfnamefont {Z.~X.}\
  \bibnamefont {{Shen}}}, \bibinfo {author} {\bibfnamefont {D.~L.}\
  \bibnamefont {{Feng}}}, \bibinfo {author} {\bibfnamefont {Z.}~\bibnamefont
  {{Hussain}}}, \ and\ \bibinfo {author} {\bibfnamefont {Y.~L.}\ \bibnamefont
  {{Chen}}},\ }\href {\doibase 10.1038/nmat3990} {\bibfield  {journal}
  {\bibinfo  {journal} {Nature Materials}\ }\textbf {\bibinfo {volume} {13}},\
  \bibinfo {pages} {677} (\bibinfo {year} {2014}{\natexlab{b}})}\BibitemShut
  {NoStop}%
\bibitem [{\citenamefont {{Steinberg}}\ \emph {et~al.}(2014)\citenamefont
  {{Steinberg}}, \citenamefont {{Young}}, \citenamefont {{Zaheer}},
  \citenamefont {{Kane}}, \citenamefont {{Mele}},\ and\ \citenamefont
  {{Rappe}}}]{Steinberg14}%
  \BibitemOpen
  \bibfield  {author} {\bibinfo {author} {\bibfnamefont {J.~A.}\ \bibnamefont
  {{Steinberg}}}, \bibinfo {author} {\bibfnamefont {S.~M.}\ \bibnamefont
  {{Young}}}, \bibinfo {author} {\bibfnamefont {S.}~\bibnamefont {{Zaheer}}},
  \bibinfo {author} {\bibfnamefont {C.~L.}\ \bibnamefont {{Kane}}}, \bibinfo
  {author} {\bibfnamefont {E.~J.}\ \bibnamefont {{Mele}}}, \ and\ \bibinfo
  {author} {\bibfnamefont {A.~M.}\ \bibnamefont {{Rappe}}},\ }\href {\doibase
  10.1103/PhysRevLett.112.036403} {\bibfield  {journal} {\bibinfo  {journal}
  {\prl}\ }\textbf {\bibinfo {volume} {112}},\ \bibinfo {eid} {036403}
  (\bibinfo {year} {2014})},\ \Eprint {http://arxiv.org/abs/1309.5967}
  {arXiv:1309.5967 [cond-mat.mtrl-sci]} \BibitemShut {NoStop}%
\bibitem [{\citenamefont {{Yang}}\ and\ \citenamefont
  {{Nagaosa}}(2014)}]{nagaosa2014}%
  \BibitemOpen
  \bibfield  {author} {\bibinfo {author} {\bibfnamefont {B.-J.}\ \bibnamefont
  {{Yang}}}\ and\ \bibinfo {author} {\bibfnamefont {N.}~\bibnamefont
  {{Nagaosa}}},\ }\href {\doibase 10.1038/ncomms5898} {\bibfield  {journal}
  {\bibinfo  {journal} {Nature Communications}\ }\textbf {\bibinfo {volume}
  {5}},\ \bibinfo {eid} {4898} (\bibinfo {year} {2014})},\ \Eprint
  {http://arxiv.org/abs/1404.0754} {arXiv:1404.0754 [cond-mat.mes-hall]}
  \BibitemShut {NoStop}%
\bibitem [{\citenamefont {{Bradlyn}}\ \emph {et~al.}(2016)\citenamefont
  {{Bradlyn}}, \citenamefont {{Cano}}, \citenamefont {{Wang}}, \citenamefont
  {{Vergniory}}, \citenamefont {{Felser}}, \citenamefont {{Cava}},\ and\
  \citenamefont {{Bernevig}}}]{bradlyn2016beyond}%
  \BibitemOpen
  \bibfield  {author} {\bibinfo {author} {\bibfnamefont {B.}~\bibnamefont
  {{Bradlyn}}}, \bibinfo {author} {\bibfnamefont {J.}~\bibnamefont {{Cano}}},
  \bibinfo {author} {\bibfnamefont {Z.}~\bibnamefont {{Wang}}}, \bibinfo
  {author} {\bibfnamefont {M.~G.}\ \bibnamefont {{Vergniory}}}, \bibinfo
  {author} {\bibfnamefont {C.}~\bibnamefont {{Felser}}}, \bibinfo {author}
  {\bibfnamefont {R.~J.}\ \bibnamefont {{Cava}}}, \ and\ \bibinfo {author}
  {\bibfnamefont {B.~A.}\ \bibnamefont {{Bernevig}}},\ }\href {\doibase
  10.1126/science.aaf5037} {\bibfield  {journal} {\bibinfo  {journal}
  {Science}\ }\textbf {\bibinfo {volume} {353}},\ \bibinfo {pages} {aaf5037}
  (\bibinfo {year} {2016})},\ \Eprint {http://arxiv.org/abs/1603.03093}
  {arXiv:1603.03093 [cond-mat.mes-hall]} \BibitemShut {NoStop}%
\bibitem [{\citenamefont {{Cano}}\ \emph {et~al.}(2019)\citenamefont {{Cano}},
  \citenamefont {{Bradlyn}},\ and\ \citenamefont
  {{Vergniory}}}]{cano2019multifold}%
  \BibitemOpen
  \bibfield  {author} {\bibinfo {author} {\bibfnamefont {J.}~\bibnamefont
  {{Cano}}}, \bibinfo {author} {\bibfnamefont {B.}~\bibnamefont {{Bradlyn}}}, \
  and\ \bibinfo {author} {\bibfnamefont {M.~G.}\ \bibnamefont {{Vergniory}}},\
  }\href {\doibase 10.1063/1.5124314} {\bibfield  {journal} {\bibinfo
  {journal} {APL Materials}\ }\textbf {\bibinfo {volume} {7}},\ \bibinfo {eid}
  {101125} (\bibinfo {year} {2019})},\ \Eprint
  {http://arxiv.org/abs/1904.12867} {arXiv:1904.12867 [cond-mat.mes-hall]}
  \BibitemShut {NoStop}%
\bibitem [{\citenamefont {{Klemenz}}\ \emph {et~al.}(2020)\citenamefont
  {{Klemenz}}, \citenamefont {{Schoop}},\ and\ \citenamefont
  {{Cano}}}]{klemenz2020systematic}%
  \BibitemOpen
  \bibfield  {author} {\bibinfo {author} {\bibfnamefont {S.}~\bibnamefont
  {{Klemenz}}}, \bibinfo {author} {\bibfnamefont {L.}~\bibnamefont {{Schoop}}},
  \ and\ \bibinfo {author} {\bibfnamefont {J.}~\bibnamefont {{Cano}}},\ }\href
  {\doibase 10.1103/PhysRevB.101.165121} {\bibfield  {journal} {\bibinfo
  {journal} {\prb}\ }\textbf {\bibinfo {volume} {101}},\ \bibinfo {eid}
  {165121} (\bibinfo {year} {2020})},\ \Eprint
  {http://arxiv.org/abs/1910.14036} {arXiv:1910.14036 [cond-mat.mes-hall]}
  \BibitemShut {NoStop}%
\bibitem [{\citenamefont {{Wieder}}\ \emph {et~al.}(2020)\citenamefont
  {{Wieder}}, \citenamefont {{Wang}}, \citenamefont {{Cano}}, \citenamefont
  {{Dai}}, \citenamefont {{Schoop}}, \citenamefont {{Bradlyn}},\ and\
  \citenamefont {{Bernevig}}}]{wieder2020strong}%
  \BibitemOpen
  \bibfield  {author} {\bibinfo {author} {\bibfnamefont {B.~J.}\ \bibnamefont
  {{Wieder}}}, \bibinfo {author} {\bibfnamefont {Z.}~\bibnamefont {{Wang}}},
  \bibinfo {author} {\bibfnamefont {J.}~\bibnamefont {{Cano}}}, \bibinfo
  {author} {\bibfnamefont {X.}~\bibnamefont {{Dai}}}, \bibinfo {author}
  {\bibfnamefont {L.~M.}\ \bibnamefont {{Schoop}}}, \bibinfo {author}
  {\bibfnamefont {B.}~\bibnamefont {{Bradlyn}}}, \ and\ \bibinfo {author}
  {\bibfnamefont {B.~A.}\ \bibnamefont {{Bernevig}}},\ }\href {\doibase
  10.1038/s41467-020-14443-5} {\bibfield  {journal} {\bibinfo  {journal}
  {Nature Communications}\ }\textbf {\bibinfo {volume} {11}},\ \bibinfo {eid}
  {627} (\bibinfo {year} {2020})},\ \Eprint {http://arxiv.org/abs/1908.00016}
  {arXiv:1908.00016 [cond-mat.mes-hall]} \BibitemShut {NoStop}%
\bibitem [{\citenamefont {{Wan}}\ \emph {et~al.}(2011)\citenamefont {{Wan}},
  \citenamefont {{Turner}}, \citenamefont {{Vishwanath}},\ and\ \citenamefont
  {{Savrasov}}}]{Wan11}%
  \BibitemOpen
  \bibfield  {author} {\bibinfo {author} {\bibfnamefont {X.}~\bibnamefont
  {{Wan}}}, \bibinfo {author} {\bibfnamefont {A.~M.}\ \bibnamefont {{Turner}}},
  \bibinfo {author} {\bibfnamefont {A.}~\bibnamefont {{Vishwanath}}}, \ and\
  \bibinfo {author} {\bibfnamefont {S.~Y.}\ \bibnamefont {{Savrasov}}},\ }\href
  {\doibase 10.1103/PhysRevB.83.205101} {\bibfield  {journal} {\bibinfo
  {journal} {\prb}\ }\textbf {\bibinfo {volume} {83}},\ \bibinfo {eid} {205101}
  (\bibinfo {year} {2011})},\ \Eprint {http://arxiv.org/abs/1007.0016}
  {arXiv:1007.0016 [cond-mat.str-el]} \BibitemShut {NoStop}%
\bibitem [{\citenamefont {{Weng}}\ \emph {et~al.}(2015)\citenamefont {{Weng}},
  \citenamefont {{Fang}}, \citenamefont {{Fang}}, \citenamefont {{Bernevig}},\
  and\ \citenamefont {{Dai}}}]{Weng15}%
  \BibitemOpen
  \bibfield  {author} {\bibinfo {author} {\bibfnamefont {H.}~\bibnamefont
  {{Weng}}}, \bibinfo {author} {\bibfnamefont {C.}~\bibnamefont {{Fang}}},
  \bibinfo {author} {\bibfnamefont {Z.}~\bibnamefont {{Fang}}}, \bibinfo
  {author} {\bibfnamefont {B.~A.}\ \bibnamefont {{Bernevig}}}, \ and\ \bibinfo
  {author} {\bibfnamefont {X.}~\bibnamefont {{Dai}}},\ }\href {\doibase
  10.1103/PhysRevX.5.011029} {\bibfield  {journal} {\bibinfo  {journal}
  {Physical Review X}\ }\textbf {\bibinfo {volume} {5}},\ \bibinfo {eid}
  {011029} (\bibinfo {year} {2015})},\ \Eprint
  {http://arxiv.org/abs/1501.00060} {arXiv:1501.00060 [cond-mat.mtrl-sci]}
  \BibitemShut {NoStop}%
\bibitem [{\citenamefont {{Huang}}\ \emph
  {et~al.}(2015{\natexlab{a}})\citenamefont {{Huang}}, \citenamefont {{Xu}},
  \citenamefont {{Belopolski}}, \citenamefont {{Lee}}, \citenamefont {{Chang}},
  \citenamefont {{Wang}}, \citenamefont {{Alidoust}}, \citenamefont {{Bian}},
  \citenamefont {{Neupane}}, \citenamefont {{Zhang}}, \citenamefont {{Jia}},
  \citenamefont {{Bansil}}, \citenamefont {{Lin}},\ and\ \citenamefont
  {{Hasan}}}]{Huang15}%
  \BibitemOpen
  \bibfield  {author} {\bibinfo {author} {\bibfnamefont {S.-M.}\ \bibnamefont
  {{Huang}}}, \bibinfo {author} {\bibfnamefont {S.-Y.}\ \bibnamefont {{Xu}}},
  \bibinfo {author} {\bibfnamefont {I.}~\bibnamefont {{Belopolski}}}, \bibinfo
  {author} {\bibfnamefont {C.-C.}\ \bibnamefont {{Lee}}}, \bibinfo {author}
  {\bibfnamefont {G.}~\bibnamefont {{Chang}}}, \bibinfo {author} {\bibfnamefont
  {B.}~\bibnamefont {{Wang}}}, \bibinfo {author} {\bibfnamefont
  {N.}~\bibnamefont {{Alidoust}}}, \bibinfo {author} {\bibfnamefont
  {G.}~\bibnamefont {{Bian}}}, \bibinfo {author} {\bibfnamefont
  {M.}~\bibnamefont {{Neupane}}}, \bibinfo {author} {\bibfnamefont
  {C.}~\bibnamefont {{Zhang}}}, \bibinfo {author} {\bibfnamefont
  {S.}~\bibnamefont {{Jia}}}, \bibinfo {author} {\bibfnamefont
  {A.}~\bibnamefont {{Bansil}}}, \bibinfo {author} {\bibfnamefont
  {H.}~\bibnamefont {{Lin}}}, \ and\ \bibinfo {author} {\bibfnamefont {M.~Z.}\
  \bibnamefont {{Hasan}}},\ }\href {\doibase 10.1038/ncomms8373} {\bibfield
  {journal} {\bibinfo  {journal} {Nature Communications}\ }\textbf {\bibinfo
  {volume} {6}},\ \bibinfo {eid} {7373} (\bibinfo {year}
  {2015}{\natexlab{a}})}\BibitemShut {NoStop}%
\bibitem [{\citenamefont {{Xu}}\ \emph
  {et~al.}(2015{\natexlab{a}})\citenamefont {{Xu}}, \citenamefont {{Alidoust}},
  \citenamefont {{Belopolski}}, \citenamefont {{Yuan}}, \citenamefont {{Bian}},
  \citenamefont {{Chang}}, \citenamefont {{Zheng}}, \citenamefont {{Strocov}},
  \citenamefont {{Sanchez}}, \citenamefont {{Chang}}, \citenamefont {{Zhang}},
  \citenamefont {{Mou}}, \citenamefont {{Wu}}, \citenamefont {{Huang}},
  \citenamefont {{Lee}}, \citenamefont {{Huang}}, \citenamefont {{Wang}},
  \citenamefont {{Bansil}}, \citenamefont {{Jeng}}, \citenamefont {{Neupert}},
  \citenamefont {{Kaminski}}, \citenamefont {{Lin}}, \citenamefont {{Jia}},\
  and\ \citenamefont {{Zahid Hasan}}}]{xu2015discovery}%
  \BibitemOpen
  \bibfield  {author} {\bibinfo {author} {\bibfnamefont {S.-Y.}\ \bibnamefont
  {{Xu}}}, \bibinfo {author} {\bibfnamefont {N.}~\bibnamefont {{Alidoust}}},
  \bibinfo {author} {\bibfnamefont {I.}~\bibnamefont {{Belopolski}}}, \bibinfo
  {author} {\bibfnamefont {Z.}~\bibnamefont {{Yuan}}}, \bibinfo {author}
  {\bibfnamefont {G.}~\bibnamefont {{Bian}}}, \bibinfo {author} {\bibfnamefont
  {T.-R.}\ \bibnamefont {{Chang}}}, \bibinfo {author} {\bibfnamefont
  {H.}~\bibnamefont {{Zheng}}}, \bibinfo {author} {\bibfnamefont {V.~N.}\
  \bibnamefont {{Strocov}}}, \bibinfo {author} {\bibfnamefont {D.~S.}\
  \bibnamefont {{Sanchez}}}, \bibinfo {author} {\bibfnamefont {G.}~\bibnamefont
  {{Chang}}}, \bibinfo {author} {\bibfnamefont {C.}~\bibnamefont {{Zhang}}},
  \bibinfo {author} {\bibfnamefont {D.}~\bibnamefont {{Mou}}}, \bibinfo
  {author} {\bibfnamefont {Y.}~\bibnamefont {{Wu}}}, \bibinfo {author}
  {\bibfnamefont {L.}~\bibnamefont {{Huang}}}, \bibinfo {author} {\bibfnamefont
  {C.-C.}\ \bibnamefont {{Lee}}}, \bibinfo {author} {\bibfnamefont {S.-M.}\
  \bibnamefont {{Huang}}}, \bibinfo {author} {\bibfnamefont {B.}~\bibnamefont
  {{Wang}}}, \bibinfo {author} {\bibfnamefont {A.}~\bibnamefont {{Bansil}}},
  \bibinfo {author} {\bibfnamefont {H.-T.}\ \bibnamefont {{Jeng}}}, \bibinfo
  {author} {\bibfnamefont {T.}~\bibnamefont {{Neupert}}}, \bibinfo {author}
  {\bibfnamefont {A.}~\bibnamefont {{Kaminski}}}, \bibinfo {author}
  {\bibfnamefont {H.}~\bibnamefont {{Lin}}}, \bibinfo {author} {\bibfnamefont
  {S.}~\bibnamefont {{Jia}}}, \ and\ \bibinfo {author} {\bibfnamefont
  {M.}~\bibnamefont {{Zahid Hasan}}},\ }\href {\doibase 10.1038/nphys3437}
  {\bibfield  {journal} {\bibinfo  {journal} {Nature Physics}\ }\textbf
  {\bibinfo {volume} {11}},\ \bibinfo {pages} {748} (\bibinfo {year}
  {2015}{\natexlab{a}})},\ \Eprint {http://arxiv.org/abs/1504.01350}
  {arXiv:1504.01350 [cond-mat.mes-hall]} \BibitemShut {NoStop}%
\bibitem [{\citenamefont {{Lv}}\ \emph
  {et~al.}(2015{\natexlab{a}})\citenamefont {{Lv}}, \citenamefont {{Xu}},
  \citenamefont {{Weng}}, \citenamefont {{Ma}}, \citenamefont {{Richard}},
  \citenamefont {{Huang}}, \citenamefont {{Zhao}}, \citenamefont {{Chen}},
  \citenamefont {{Matt}}, \citenamefont {{Bisti}}, \citenamefont {{Strocov}},
  \citenamefont {{Mesot}}, \citenamefont {{Fang}}, \citenamefont {{Dai}},
  \citenamefont {{Qian}}, \citenamefont {{Shi}},\ and\ \citenamefont
  {{Ding}}}]{lv2015Nat}%
  \BibitemOpen
  \bibfield  {author} {\bibinfo {author} {\bibfnamefont {B.~Q.}\ \bibnamefont
  {{Lv}}}, \bibinfo {author} {\bibfnamefont {N.}~\bibnamefont {{Xu}}}, \bibinfo
  {author} {\bibfnamefont {H.~M.}\ \bibnamefont {{Weng}}}, \bibinfo {author}
  {\bibfnamefont {J.~Z.}\ \bibnamefont {{Ma}}}, \bibinfo {author}
  {\bibfnamefont {P.}~\bibnamefont {{Richard}}}, \bibinfo {author}
  {\bibfnamefont {X.~C.}\ \bibnamefont {{Huang}}}, \bibinfo {author}
  {\bibfnamefont {L.~X.}\ \bibnamefont {{Zhao}}}, \bibinfo {author}
  {\bibfnamefont {G.~F.}\ \bibnamefont {{Chen}}}, \bibinfo {author}
  {\bibfnamefont {C.~E.}\ \bibnamefont {{Matt}}}, \bibinfo {author}
  {\bibfnamefont {F.}~\bibnamefont {{Bisti}}}, \bibinfo {author} {\bibfnamefont
  {V.~N.}\ \bibnamefont {{Strocov}}}, \bibinfo {author} {\bibfnamefont
  {J.}~\bibnamefont {{Mesot}}}, \bibinfo {author} {\bibfnamefont
  {Z.}~\bibnamefont {{Fang}}}, \bibinfo {author} {\bibfnamefont
  {X.}~\bibnamefont {{Dai}}}, \bibinfo {author} {\bibfnamefont
  {T.}~\bibnamefont {{Qian}}}, \bibinfo {author} {\bibfnamefont
  {M.}~\bibnamefont {{Shi}}}, \ and\ \bibinfo {author} {\bibfnamefont
  {H.}~\bibnamefont {{Ding}}},\ }\href {\doibase 10.1038/nphys3426} {\bibfield
  {journal} {\bibinfo  {journal} {Nature Physics}\ }\textbf {\bibinfo {volume}
  {11}},\ \bibinfo {pages} {724} (\bibinfo {year} {2015}{\natexlab{a}})},\
  \Eprint {http://arxiv.org/abs/1503.09188} {arXiv:1503.09188
  [cond-mat.mtrl-sci]} \BibitemShut {NoStop}%
\bibitem [{\citenamefont {{Xu}}\ \emph
  {et~al.}(2015{\natexlab{b}})\citenamefont {{Xu}}, \citenamefont
  {{Belopolski}}, \citenamefont {{Alidoust}}, \citenamefont {{Neupane}},
  \citenamefont {{Bian}}, \citenamefont {{Zhang}}, \citenamefont {{Sankar}},
  \citenamefont {{Chang}}, \citenamefont {{Yuan}}, \citenamefont {{Lee}},
  \citenamefont {{Huang}}, \citenamefont {{Zheng}}, \citenamefont {{Ma}},
  \citenamefont {{Sanchez}}, \citenamefont {{Wang}}, \citenamefont {{Bansil}},
  \citenamefont {{Chou}}, \citenamefont {{Shibayev}}, \citenamefont {{Lin}},
  \citenamefont {{Jia}},\ and\ \citenamefont {{Hasan}}}]{xu2015}%
  \BibitemOpen
  \bibfield  {author} {\bibinfo {author} {\bibfnamefont {S.-Y.}\ \bibnamefont
  {{Xu}}}, \bibinfo {author} {\bibfnamefont {I.}~\bibnamefont {{Belopolski}}},
  \bibinfo {author} {\bibfnamefont {N.}~\bibnamefont {{Alidoust}}}, \bibinfo
  {author} {\bibfnamefont {M.}~\bibnamefont {{Neupane}}}, \bibinfo {author}
  {\bibfnamefont {G.}~\bibnamefont {{Bian}}}, \bibinfo {author} {\bibfnamefont
  {C.}~\bibnamefont {{Zhang}}}, \bibinfo {author} {\bibfnamefont
  {R.}~\bibnamefont {{Sankar}}}, \bibinfo {author} {\bibfnamefont
  {G.}~\bibnamefont {{Chang}}}, \bibinfo {author} {\bibfnamefont
  {Z.}~\bibnamefont {{Yuan}}}, \bibinfo {author} {\bibfnamefont {C.-C.}\
  \bibnamefont {{Lee}}}, \bibinfo {author} {\bibfnamefont {S.-M.}\ \bibnamefont
  {{Huang}}}, \bibinfo {author} {\bibfnamefont {H.}~\bibnamefont {{Zheng}}},
  \bibinfo {author} {\bibfnamefont {J.}~\bibnamefont {{Ma}}}, \bibinfo {author}
  {\bibfnamefont {D.~S.}\ \bibnamefont {{Sanchez}}}, \bibinfo {author}
  {\bibfnamefont {B.}~\bibnamefont {{Wang}}}, \bibinfo {author} {\bibfnamefont
  {A.}~\bibnamefont {{Bansil}}}, \bibinfo {author} {\bibfnamefont
  {F.}~\bibnamefont {{Chou}}}, \bibinfo {author} {\bibfnamefont {P.~P.}\
  \bibnamefont {{Shibayev}}}, \bibinfo {author} {\bibfnamefont
  {H.}~\bibnamefont {{Lin}}}, \bibinfo {author} {\bibfnamefont
  {S.}~\bibnamefont {{Jia}}}, \ and\ \bibinfo {author} {\bibfnamefont {M.~Z.}\
  \bibnamefont {{Hasan}}},\ }\href {\doibase 10.1126/science.aaa9297}
  {\bibfield  {journal} {\bibinfo  {journal} {Science}\ }\textbf {\bibinfo
  {volume} {349}},\ \bibinfo {pages} {613} (\bibinfo {year}
  {2015}{\natexlab{b}})},\ \Eprint {http://arxiv.org/abs/1502.03807}
  {arXiv:1502.03807 [cond-mat.mes-hall]} \BibitemShut {NoStop}%
\bibitem [{\citenamefont {{Lv}}\ \emph
  {et~al.}(2015{\natexlab{b}})\citenamefont {{Lv}}, \citenamefont {{Weng}},
  \citenamefont {{Fu}}, \citenamefont {{Wang}}, \citenamefont {{Miao}},
  \citenamefont {{Ma}}, \citenamefont {{Richard}}, \citenamefont {{Huang}},
  \citenamefont {{Zhao}}, \citenamefont {{Chen}}, \citenamefont {{Fang}},
  \citenamefont {{Dai}}, \citenamefont {{Qian}},\ and\ \citenamefont
  {{Ding}}}]{lv2015PRX}%
  \BibitemOpen
  \bibfield  {author} {\bibinfo {author} {\bibfnamefont {B.~Q.}\ \bibnamefont
  {{Lv}}}, \bibinfo {author} {\bibfnamefont {H.~M.}\ \bibnamefont {{Weng}}},
  \bibinfo {author} {\bibfnamefont {B.~B.}\ \bibnamefont {{Fu}}}, \bibinfo
  {author} {\bibfnamefont {X.~P.}\ \bibnamefont {{Wang}}}, \bibinfo {author}
  {\bibfnamefont {H.}~\bibnamefont {{Miao}}}, \bibinfo {author} {\bibfnamefont
  {J.}~\bibnamefont {{Ma}}}, \bibinfo {author} {\bibfnamefont {P.}~\bibnamefont
  {{Richard}}}, \bibinfo {author} {\bibfnamefont {X.~C.}\ \bibnamefont
  {{Huang}}}, \bibinfo {author} {\bibfnamefont {L.~X.}\ \bibnamefont {{Zhao}}},
  \bibinfo {author} {\bibfnamefont {G.~F.}\ \bibnamefont {{Chen}}}, \bibinfo
  {author} {\bibfnamefont {Z.}~\bibnamefont {{Fang}}}, \bibinfo {author}
  {\bibfnamefont {X.}~\bibnamefont {{Dai}}}, \bibinfo {author} {\bibfnamefont
  {T.}~\bibnamefont {{Qian}}}, \ and\ \bibinfo {author} {\bibfnamefont
  {H.}~\bibnamefont {{Ding}}},\ }\href {\doibase 10.1103/PhysRevX.5.031013}
  {\bibfield  {journal} {\bibinfo  {journal} {Physical Review X}\ }\textbf
  {\bibinfo {volume} {5}},\ \bibinfo {eid} {031013} (\bibinfo {year}
  {2015}{\natexlab{b}})},\ \Eprint {http://arxiv.org/abs/1502.04684}
  {arXiv:1502.04684 [cond-mat.mtrl-sci]} \BibitemShut {NoStop}%
\bibitem [{\citenamefont {{Xiong}}\ \emph {et~al.}(2015)\citenamefont
  {{Xiong}}, \citenamefont {{Kushwaha}}, \citenamefont {{Liang}}, \citenamefont
  {{Krizan}}, \citenamefont {{Hirschberger}}, \citenamefont {{Wang}},
  \citenamefont {{Cava}},\ and\ \citenamefont {{Ong}}}]{2015Xiong}%
  \BibitemOpen
  \bibfield  {author} {\bibinfo {author} {\bibfnamefont {J.}~\bibnamefont
  {{Xiong}}}, \bibinfo {author} {\bibfnamefont {S.~K.}\ \bibnamefont
  {{Kushwaha}}}, \bibinfo {author} {\bibfnamefont {T.}~\bibnamefont {{Liang}}},
  \bibinfo {author} {\bibfnamefont {J.~W.}\ \bibnamefont {{Krizan}}}, \bibinfo
  {author} {\bibfnamefont {M.}~\bibnamefont {{Hirschberger}}}, \bibinfo
  {author} {\bibfnamefont {W.}~\bibnamefont {{Wang}}}, \bibinfo {author}
  {\bibfnamefont {R.~J.}\ \bibnamefont {{Cava}}}, \ and\ \bibinfo {author}
  {\bibfnamefont {N.~P.}\ \bibnamefont {{Ong}}},\ }\href {\doibase
  10.1126/science.aac6089} {\bibfield  {journal} {\bibinfo  {journal}
  {Science}\ }\textbf {\bibinfo {volume} {350}},\ \bibinfo {pages} {413}
  (\bibinfo {year} {2015})}\BibitemShut {NoStop}%
\bibitem [{\citenamefont {{Armitage}}\ \emph {et~al.}(2018)\citenamefont
  {{Armitage}}, \citenamefont {{Mele}},\ and\ \citenamefont
  {{Vishwanath}}}]{armitage2018}%
  \BibitemOpen
  \bibfield  {author} {\bibinfo {author} {\bibfnamefont {N.~P.}\ \bibnamefont
  {{Armitage}}}, \bibinfo {author} {\bibfnamefont {E.~J.}\ \bibnamefont
  {{Mele}}}, \ and\ \bibinfo {author} {\bibfnamefont {A.}~\bibnamefont
  {{Vishwanath}}},\ }\href {\doibase 10.1103/RevModPhys.90.015001} {\bibfield
  {journal} {\bibinfo  {journal} {Reviews of Modern Physics}\ }\textbf
  {\bibinfo {volume} {90}},\ \bibinfo {eid} {015001} (\bibinfo {year}
  {2018})},\ \Eprint {http://arxiv.org/abs/1705.01111} {arXiv:1705.01111
  [cond-mat.str-el]} \BibitemShut {NoStop}%
\bibitem [{\citenamefont {{Fukushima}}\ \emph {et~al.}(2008)\citenamefont
  {{Fukushima}}, \citenamefont {{Kharzeev}},\ and\ \citenamefont
  {{Warringa}}}]{2008Fukushima}%
  \BibitemOpen
  \bibfield  {author} {\bibinfo {author} {\bibfnamefont {K.}~\bibnamefont
  {{Fukushima}}}, \bibinfo {author} {\bibfnamefont {D.~E.}\ \bibnamefont
  {{Kharzeev}}}, \ and\ \bibinfo {author} {\bibfnamefont {H.~J.}\ \bibnamefont
  {{Warringa}}},\ }\href {\doibase 10.1103/PhysRevD.78.074033} {\bibfield
  {journal} {\bibinfo  {journal} {\prd}\ }\textbf {\bibinfo {volume} {78}},\
  \bibinfo {eid} {074033} (\bibinfo {year} {2008})},\ \Eprint
  {http://arxiv.org/abs/0808.3382} {arXiv:0808.3382 [hep-ph]} \BibitemShut
  {NoStop}%
\bibitem [{\citenamefont {{Aji}}(2012)}]{aji2012adler}%
  \BibitemOpen
  \bibfield  {author} {\bibinfo {author} {\bibfnamefont {V.}~\bibnamefont
  {{Aji}}},\ }\href {\doibase 10.1103/PhysRevB.85.241101} {\bibfield  {journal}
  {\bibinfo  {journal} {\prb}\ }\textbf {\bibinfo {volume} {85}},\ \bibinfo
  {eid} {241101} (\bibinfo {year} {2012})},\ \Eprint
  {http://arxiv.org/abs/1108.4426} {arXiv:1108.4426 [cond-mat.str-el]}
  \BibitemShut {NoStop}%
\bibitem [{\citenamefont {{Son}}\ and\ \citenamefont
  {{Spivak}}(2013)}]{2013Son}%
  \BibitemOpen
  \bibfield  {author} {\bibinfo {author} {\bibfnamefont {D.~T.}\ \bibnamefont
  {{Son}}}\ and\ \bibinfo {author} {\bibfnamefont {B.~Z.}\ \bibnamefont
  {{Spivak}}},\ }\href {\doibase 10.1103/PhysRevB.88.104412} {\bibfield
  {journal} {\bibinfo  {journal} {\prb}\ }\textbf {\bibinfo {volume} {88}},\
  \bibinfo {eid} {104412} (\bibinfo {year} {2013})},\ \Eprint
  {http://arxiv.org/abs/1206.1627} {arXiv:1206.1627 [cond-mat.mes-hall]}
  \BibitemShut {NoStop}%
\bibitem [{\citenamefont {{Burkov}}(2014)}]{2014Burkov}%
  \BibitemOpen
  \bibfield  {author} {\bibinfo {author} {\bibfnamefont {A.~A.}\ \bibnamefont
  {{Burkov}}},\ }\href {\doibase 10.1103/PhysRevLett.113.247203} {\bibfield
  {journal} {\bibinfo  {journal} {\prl}\ }\textbf {\bibinfo {volume} {113}},\
  \bibinfo {eid} {247203} (\bibinfo {year} {2014})},\ \Eprint
  {http://arxiv.org/abs/1409.0013} {arXiv:1409.0013 [cond-mat.mes-hall]}
  \BibitemShut {NoStop}%
\bibitem [{\citenamefont {{Li}}\ \emph {et~al.}(2016)\citenamefont {{Li}},
  \citenamefont {{Kharzeev}}, \citenamefont {{Zhang}}, \citenamefont {{Huang}},
  \citenamefont {{Pletikosi{\'c}}}, \citenamefont {{Fedorov}}, \citenamefont
  {{Zhong}}, \citenamefont {{Schneeloch}}, \citenamefont {{Gu}},\ and\
  \citenamefont {{Valla}}}]{2016Li}%
  \BibitemOpen
  \bibfield  {author} {\bibinfo {author} {\bibfnamefont {Q.}~\bibnamefont
  {{Li}}}, \bibinfo {author} {\bibfnamefont {D.~E.}\ \bibnamefont
  {{Kharzeev}}}, \bibinfo {author} {\bibfnamefont {C.}~\bibnamefont {{Zhang}}},
  \bibinfo {author} {\bibfnamefont {Y.}~\bibnamefont {{Huang}}}, \bibinfo
  {author} {\bibfnamefont {I.}~\bibnamefont {{Pletikosi{\'c}}}}, \bibinfo
  {author} {\bibfnamefont {A.~V.}\ \bibnamefont {{Fedorov}}}, \bibinfo {author}
  {\bibfnamefont {R.~D.}\ \bibnamefont {{Zhong}}}, \bibinfo {author}
  {\bibfnamefont {J.~A.}\ \bibnamefont {{Schneeloch}}}, \bibinfo {author}
  {\bibfnamefont {G.~D.}\ \bibnamefont {{Gu}}}, \ and\ \bibinfo {author}
  {\bibfnamefont {T.}~\bibnamefont {{Valla}}},\ }\href {\doibase
  10.1038/nphys3648} {\bibfield  {journal} {\bibinfo  {journal} {Nature
  Physics}\ }\textbf {\bibinfo {volume} {12}},\ \bibinfo {pages} {550}
  (\bibinfo {year} {2016})},\ \Eprint {http://arxiv.org/abs/1412.6543}
  {arXiv:1412.6543 [cond-mat.str-el]} \BibitemShut {NoStop}%
\bibitem [{\citenamefont {{Huang}}\ \emph
  {et~al.}(2015{\natexlab{b}})\citenamefont {{Huang}}, \citenamefont {{Zhao}},
  \citenamefont {{Long}}, \citenamefont {{Wang}}, \citenamefont {{Chen}},
  \citenamefont {{Yang}}, \citenamefont {{Liang}}, \citenamefont {{Xue}},
  \citenamefont {{Weng}}, \citenamefont {{Fang}}, \citenamefont {{Dai}},\ and\
  \citenamefont {{Chen}}}]{Huang2015}%
  \BibitemOpen
  \bibfield  {author} {\bibinfo {author} {\bibfnamefont {X.}~\bibnamefont
  {{Huang}}}, \bibinfo {author} {\bibfnamefont {L.}~\bibnamefont {{Zhao}}},
  \bibinfo {author} {\bibfnamefont {Y.}~\bibnamefont {{Long}}}, \bibinfo
  {author} {\bibfnamefont {P.}~\bibnamefont {{Wang}}}, \bibinfo {author}
  {\bibfnamefont {D.}~\bibnamefont {{Chen}}}, \bibinfo {author} {\bibfnamefont
  {Z.}~\bibnamefont {{Yang}}}, \bibinfo {author} {\bibfnamefont
  {H.}~\bibnamefont {{Liang}}}, \bibinfo {author} {\bibfnamefont
  {M.}~\bibnamefont {{Xue}}}, \bibinfo {author} {\bibfnamefont
  {H.}~\bibnamefont {{Weng}}}, \bibinfo {author} {\bibfnamefont
  {Z.}~\bibnamefont {{Fang}}}, \bibinfo {author} {\bibfnamefont
  {X.}~\bibnamefont {{Dai}}}, \ and\ \bibinfo {author} {\bibfnamefont
  {G.}~\bibnamefont {{Chen}}},\ }\href {\doibase 10.1103/PhysRevX.5.031023}
  {\bibfield  {journal} {\bibinfo  {journal} {Physical Review X}\ }\textbf
  {\bibinfo {volume} {5}},\ \bibinfo {eid} {031023} (\bibinfo {year}
  {2015}{\natexlab{b}})},\ \Eprint {http://arxiv.org/abs/1503.01304}
  {arXiv:1503.01304 [cond-mat.mtrl-sci]} \BibitemShut {NoStop}%
\bibitem [{\citenamefont {{Zhang}}\ \emph {et~al.}(2016)\citenamefont
  {{Zhang}}, \citenamefont {{Xu}}, \citenamefont {{Belopolski}}, \citenamefont
  {{Yuan}}, \citenamefont {{Lin}}, \citenamefont {{Tong}}, \citenamefont
  {{Bian}}, \citenamefont {{Alidoust}}, \citenamefont {{Lee}}, \citenamefont
  {{Huang}}, \citenamefont {{Chang}}, \citenamefont {{Chang}}, \citenamefont
  {{Hsu}}, \citenamefont {{Jeng}}, \citenamefont {{Neupane}}, \citenamefont
  {{Sanchez}}, \citenamefont {{Zheng}}, \citenamefont {{Wang}}, \citenamefont
  {{Lin}}, \citenamefont {{Zhang}}, \citenamefont {{Lu}}, \citenamefont
  {{Shen}}, \citenamefont {{Neupert}}, \citenamefont {{Zahid Hasan}},\ and\
  \citenamefont {{Jia}}}]{Zhang2016}%
  \BibitemOpen
  \bibfield  {author} {\bibinfo {author} {\bibfnamefont {C.-L.}\ \bibnamefont
  {{Zhang}}}, \bibinfo {author} {\bibfnamefont {S.-Y.}\ \bibnamefont {{Xu}}},
  \bibinfo {author} {\bibfnamefont {I.}~\bibnamefont {{Belopolski}}}, \bibinfo
  {author} {\bibfnamefont {Z.}~\bibnamefont {{Yuan}}}, \bibinfo {author}
  {\bibfnamefont {Z.}~\bibnamefont {{Lin}}}, \bibinfo {author} {\bibfnamefont
  {B.}~\bibnamefont {{Tong}}}, \bibinfo {author} {\bibfnamefont
  {G.}~\bibnamefont {{Bian}}}, \bibinfo {author} {\bibfnamefont
  {N.}~\bibnamefont {{Alidoust}}}, \bibinfo {author} {\bibfnamefont {C.-C.}\
  \bibnamefont {{Lee}}}, \bibinfo {author} {\bibfnamefont {S.-M.}\ \bibnamefont
  {{Huang}}}, \bibinfo {author} {\bibfnamefont {T.-R.}\ \bibnamefont
  {{Chang}}}, \bibinfo {author} {\bibfnamefont {G.}~\bibnamefont {{Chang}}},
  \bibinfo {author} {\bibfnamefont {C.-H.}\ \bibnamefont {{Hsu}}}, \bibinfo
  {author} {\bibfnamefont {H.-T.}\ \bibnamefont {{Jeng}}}, \bibinfo {author}
  {\bibfnamefont {M.}~\bibnamefont {{Neupane}}}, \bibinfo {author}
  {\bibfnamefont {D.~S.}\ \bibnamefont {{Sanchez}}}, \bibinfo {author}
  {\bibfnamefont {H.}~\bibnamefont {{Zheng}}}, \bibinfo {author} {\bibfnamefont
  {J.}~\bibnamefont {{Wang}}}, \bibinfo {author} {\bibfnamefont
  {H.}~\bibnamefont {{Lin}}}, \bibinfo {author} {\bibfnamefont
  {C.}~\bibnamefont {{Zhang}}}, \bibinfo {author} {\bibfnamefont {H.-Z.}\
  \bibnamefont {{Lu}}}, \bibinfo {author} {\bibfnamefont {S.-Q.}\ \bibnamefont
  {{Shen}}}, \bibinfo {author} {\bibfnamefont {T.}~\bibnamefont {{Neupert}}},
  \bibinfo {author} {\bibfnamefont {M.}~\bibnamefont {{Zahid Hasan}}}, \ and\
  \bibinfo {author} {\bibfnamefont {S.}~\bibnamefont {{Jia}}},\ }\href
  {\doibase 10.1038/ncomms10735} {\bibfield  {journal} {\bibinfo  {journal}
  {Nature Communications}\ }\textbf {\bibinfo {volume} {7}},\ \bibinfo {eid}
  {10735} (\bibinfo {year} {2016})},\ \Eprint {http://arxiv.org/abs/1601.04208}
  {arXiv:1601.04208 [cond-mat.mtrl-sci]} \BibitemShut {NoStop}%
\bibitem [{\citenamefont {{Wang}}\ \emph {et~al.}(2016)\citenamefont {{Wang}},
  \citenamefont {{Zheng}}, \citenamefont {{Shen}}, \citenamefont {{Lu}},
  \citenamefont {{Fang}}, \citenamefont {{Sheng}}, \citenamefont {{Zhou}},
  \citenamefont {{Yang}}, \citenamefont {{Li}}, \citenamefont {{Feng}},\ and\
  \citenamefont {{Xu}}}]{Wang2016}%
  \BibitemOpen
  \bibfield  {author} {\bibinfo {author} {\bibfnamefont {Z.}~\bibnamefont
  {{Wang}}}, \bibinfo {author} {\bibfnamefont {Y.}~\bibnamefont {{Zheng}}},
  \bibinfo {author} {\bibfnamefont {Z.}~\bibnamefont {{Shen}}}, \bibinfo
  {author} {\bibfnamefont {Y.}~\bibnamefont {{Lu}}}, \bibinfo {author}
  {\bibfnamefont {H.}~\bibnamefont {{Fang}}}, \bibinfo {author} {\bibfnamefont
  {F.}~\bibnamefont {{Sheng}}}, \bibinfo {author} {\bibfnamefont
  {Y.}~\bibnamefont {{Zhou}}}, \bibinfo {author} {\bibfnamefont
  {X.}~\bibnamefont {{Yang}}}, \bibinfo {author} {\bibfnamefont
  {Y.}~\bibnamefont {{Li}}}, \bibinfo {author} {\bibfnamefont {C.}~\bibnamefont
  {{Feng}}}, \ and\ \bibinfo {author} {\bibfnamefont {Z.-A.}\ \bibnamefont
  {{Xu}}},\ }\href {\doibase 10.1103/PhysRevB.93.121112} {\bibfield  {journal}
  {\bibinfo  {journal} {\prb}\ }\textbf {\bibinfo {volume} {93}},\ \bibinfo
  {eid} {121112} (\bibinfo {year} {2016})},\ \Eprint
  {http://arxiv.org/abs/1506.00924} {arXiv:1506.00924 [cond-mat.mes-hall]}
  \BibitemShut {NoStop}%
\bibitem [{\citenamefont {{Arnold}}\ \emph {et~al.}(2016)\citenamefont
  {{Arnold}}, \citenamefont {{Shekhar}}, \citenamefont {{Wu}}, \citenamefont
  {{Sun}}, \citenamefont {{Dos Reis}}, \citenamefont {{Kumar}}, \citenamefont
  {{Naumann}}, \citenamefont {{Ajeesh}}, \citenamefont {{Schmidt}},
  \citenamefont {{Grushin}}, \citenamefont {{Bardarson}}, \citenamefont
  {{Baenitz}}, \citenamefont {{Sokolov}}, \citenamefont {{Borrmann}},
  \citenamefont {{Nicklas}}, \citenamefont {{Felser}}, \citenamefont
  {{Hassinger}},\ and\ \citenamefont {{Yan}}}]{Arnold2016}%
  \BibitemOpen
  \bibfield  {author} {\bibinfo {author} {\bibfnamefont {F.}~\bibnamefont
  {{Arnold}}}, \bibinfo {author} {\bibfnamefont {C.}~\bibnamefont {{Shekhar}}},
  \bibinfo {author} {\bibfnamefont {S.-C.}\ \bibnamefont {{Wu}}}, \bibinfo
  {author} {\bibfnamefont {Y.}~\bibnamefont {{Sun}}}, \bibinfo {author}
  {\bibfnamefont {R.~D.}\ \bibnamefont {{Dos Reis}}}, \bibinfo {author}
  {\bibfnamefont {N.}~\bibnamefont {{Kumar}}}, \bibinfo {author} {\bibfnamefont
  {M.}~\bibnamefont {{Naumann}}}, \bibinfo {author} {\bibfnamefont {M.~O.}\
  \bibnamefont {{Ajeesh}}}, \bibinfo {author} {\bibfnamefont {M.}~\bibnamefont
  {{Schmidt}}}, \bibinfo {author} {\bibfnamefont {A.~G.}\ \bibnamefont
  {{Grushin}}}, \bibinfo {author} {\bibfnamefont {J.~H.}\ \bibnamefont
  {{Bardarson}}}, \bibinfo {author} {\bibfnamefont {M.}~\bibnamefont
  {{Baenitz}}}, \bibinfo {author} {\bibfnamefont {D.}~\bibnamefont
  {{Sokolov}}}, \bibinfo {author} {\bibfnamefont {H.}~\bibnamefont
  {{Borrmann}}}, \bibinfo {author} {\bibfnamefont {M.}~\bibnamefont
  {{Nicklas}}}, \bibinfo {author} {\bibfnamefont {C.}~\bibnamefont {{Felser}}},
  \bibinfo {author} {\bibfnamefont {E.}~\bibnamefont {{Hassinger}}}, \ and\
  \bibinfo {author} {\bibfnamefont {B.}~\bibnamefont {{Yan}}},\ }\href
  {\doibase 10.1038/ncomms11615} {\bibfield  {journal} {\bibinfo  {journal}
  {Nature Communications}\ }\textbf {\bibinfo {volume} {7}},\ \bibinfo {eid}
  {11615} (\bibinfo {year} {2016})},\ \Eprint {http://arxiv.org/abs/1506.06577}
  {arXiv:1506.06577 [cond-mat.mtrl-sci]} \BibitemShut {NoStop}%
\bibitem [{\citenamefont {{Chan}}\ \emph {et~al.}(2017)\citenamefont {{Chan}},
  \citenamefont {{Lindner}}, \citenamefont {{Refael}},\ and\ \citenamefont
  {{Lee}}}]{lee2017}%
  \BibitemOpen
  \bibfield  {author} {\bibinfo {author} {\bibfnamefont {C.-K.}\ \bibnamefont
  {{Chan}}}, \bibinfo {author} {\bibfnamefont {N.~H.}\ \bibnamefont
  {{Lindner}}}, \bibinfo {author} {\bibfnamefont {G.}~\bibnamefont {{Refael}}},
  \ and\ \bibinfo {author} {\bibfnamefont {P.~A.}\ \bibnamefont {{Lee}}},\
  }\href {\doibase 10.1103/PhysRevB.95.041104} {\bibfield  {journal} {\bibinfo
  {journal} {\prb}\ }\textbf {\bibinfo {volume} {95}},\ \bibinfo {eid} {041104}
  (\bibinfo {year} {2017})},\ \Eprint {http://arxiv.org/abs/1607.07839}
  {arXiv:1607.07839 [cond-mat.mes-hall]} \BibitemShut {NoStop}%
\bibitem [{\citenamefont {{Ma}}\ \emph {et~al.}(2017)\citenamefont {{Ma}},
  \citenamefont {{Xu}}, \citenamefont {{Chan}}, \citenamefont {{Zhang}},
  \citenamefont {{Chang}}, \citenamefont {{Lin}}, \citenamefont {{Xie}},
  \citenamefont {{Palacios}}, \citenamefont {{Lin}}, \citenamefont {{Jia}},
  \citenamefont {{Lee}}, \citenamefont {{Jarillo-Herrero}},\ and\ \citenamefont
  {{Gedik}}}]{ma2017}%
  \BibitemOpen
  \bibfield  {author} {\bibinfo {author} {\bibfnamefont {Q.}~\bibnamefont
  {{Ma}}}, \bibinfo {author} {\bibfnamefont {S.-Y.}\ \bibnamefont {{Xu}}},
  \bibinfo {author} {\bibfnamefont {C.-K.}\ \bibnamefont {{Chan}}}, \bibinfo
  {author} {\bibfnamefont {C.-L.}\ \bibnamefont {{Zhang}}}, \bibinfo {author}
  {\bibfnamefont {G.}~\bibnamefont {{Chang}}}, \bibinfo {author} {\bibfnamefont
  {Y.}~\bibnamefont {{Lin}}}, \bibinfo {author} {\bibfnamefont
  {W.}~\bibnamefont {{Xie}}}, \bibinfo {author} {\bibfnamefont
  {T.}~\bibnamefont {{Palacios}}}, \bibinfo {author} {\bibfnamefont
  {H.}~\bibnamefont {{Lin}}}, \bibinfo {author} {\bibfnamefont
  {S.}~\bibnamefont {{Jia}}}, \bibinfo {author} {\bibfnamefont {P.~A.}\
  \bibnamefont {{Lee}}}, \bibinfo {author} {\bibfnamefont {P.}~\bibnamefont
  {{Jarillo-Herrero}}}, \ and\ \bibinfo {author} {\bibfnamefont
  {N.}~\bibnamefont {{Gedik}}},\ }\href {\doibase 10.1038/nphys4146} {\bibfield
   {journal} {\bibinfo  {journal} {Nature Physics}\ }\textbf {\bibinfo {volume}
  {13}},\ \bibinfo {pages} {842} (\bibinfo {year} {2017})},\ \Eprint
  {http://arxiv.org/abs/1705.00590} {arXiv:1705.00590 [cond-mat.mtrl-sci]}
  \BibitemShut {NoStop}%
\bibitem [{\citenamefont {{Gao}}\ \emph {et~al.}(2020)\citenamefont {{Gao}},
  \citenamefont {{Kaushik}}, \citenamefont {{Philip}}, \citenamefont {{Li}},
  \citenamefont {{Qin}}, \citenamefont {{Liu}}, \citenamefont {{Zhang}},
  \citenamefont {{Su}}, \citenamefont {{Chen}}, \citenamefont {{Weng}},
  \citenamefont {{Kharzeev}}, \citenamefont {{Liu}},\ and\ \citenamefont
  {{Qi}}}]{gao2020chiral}%
  \BibitemOpen
  \bibfield  {author} {\bibinfo {author} {\bibfnamefont {Y.}~\bibnamefont
  {{Gao}}}, \bibinfo {author} {\bibfnamefont {S.}~\bibnamefont {{Kaushik}}},
  \bibinfo {author} {\bibfnamefont {E.~J.}\ \bibnamefont {{Philip}}}, \bibinfo
  {author} {\bibfnamefont {Z.}~\bibnamefont {{Li}}}, \bibinfo {author}
  {\bibfnamefont {Y.}~\bibnamefont {{Qin}}}, \bibinfo {author} {\bibfnamefont
  {Y.~P.}\ \bibnamefont {{Liu}}}, \bibinfo {author} {\bibfnamefont {W.~L.}\
  \bibnamefont {{Zhang}}}, \bibinfo {author} {\bibfnamefont {Y.~L.}\
  \bibnamefont {{Su}}}, \bibinfo {author} {\bibfnamefont {X.}~\bibnamefont
  {{Chen}}}, \bibinfo {author} {\bibfnamefont {H.}~\bibnamefont {{Weng}}},
  \bibinfo {author} {\bibfnamefont {D.~E.}\ \bibnamefont {{Kharzeev}}},
  \bibinfo {author} {\bibfnamefont {M.~K.}\ \bibnamefont {{Liu}}}, \ and\
  \bibinfo {author} {\bibfnamefont {J.}~\bibnamefont {{Qi}}},\ }\href {\doibase
  10.1038/s41467-020-14463-1} {\bibfield  {journal} {\bibinfo  {journal}
  {Nature Communications}\ }\textbf {\bibinfo {volume} {11}},\ \bibinfo {eid}
  {720} (\bibinfo {year} {2020})},\ \Eprint {http://arxiv.org/abs/1901.00986}
  {arXiv:1901.00986 [cond-mat.mtrl-sci]} \BibitemShut {NoStop}%
\bibitem [{\citenamefont {{de Juan}}\ \emph {et~al.}(2017)\citenamefont {{de
  Juan}}, \citenamefont {{Grushin}}, \citenamefont {{Morimoto}},\ and\
  \citenamefont {{Moore}}}]{dejuan2017}%
  \BibitemOpen
  \bibfield  {author} {\bibinfo {author} {\bibfnamefont {F.}~\bibnamefont {{de
  Juan}}}, \bibinfo {author} {\bibfnamefont {A.~G.}\ \bibnamefont {{Grushin}}},
  \bibinfo {author} {\bibfnamefont {T.}~\bibnamefont {{Morimoto}}}, \ and\
  \bibinfo {author} {\bibfnamefont {J.~E.}\ \bibnamefont {{Moore}}},\ }\href
  {\doibase 10.1038/ncomms15995} {\bibfield  {journal} {\bibinfo  {journal}
  {Nature Communications}\ }\textbf {\bibinfo {volume} {8}},\ \bibinfo {eid}
  {15995} (\bibinfo {year} {2017})},\ \Eprint {http://arxiv.org/abs/1611.05887}
  {arXiv:1611.05887 [cond-mat.str-el]} \BibitemShut {NoStop}%
\bibitem [{\citenamefont {{Kharzeev}}\ \emph
  {et~al.}(2018{\natexlab{a}})\citenamefont {{Kharzeev}}, \citenamefont
  {{Kikuchi}}, \citenamefont {{Meyer}},\ and\ \citenamefont
  {{Tanizaki}}}]{yuta2018}%
  \BibitemOpen
  \bibfield  {author} {\bibinfo {author} {\bibfnamefont {D.~E.}\ \bibnamefont
  {{Kharzeev}}}, \bibinfo {author} {\bibfnamefont {Y.}~\bibnamefont
  {{Kikuchi}}}, \bibinfo {author} {\bibfnamefont {R.}~\bibnamefont {{Meyer}}},
  \ and\ \bibinfo {author} {\bibfnamefont {Y.}~\bibnamefont {{Tanizaki}}},\
  }\href {\doibase 10.1103/PhysRevB.98.014305} {\bibfield  {journal} {\bibinfo
  {journal} {\prb}\ }\textbf {\bibinfo {volume} {98}},\ \bibinfo {eid} {014305}
  (\bibinfo {year} {2018}{\natexlab{a}})},\ \Eprint
  {http://arxiv.org/abs/1801.10283} {arXiv:1801.10283 [cond-mat.mes-hall]}
  \BibitemShut {NoStop}%
\bibitem [{\citenamefont {{Kharzeev}}\ \emph
  {et~al.}(2018{\natexlab{b}})\citenamefont {{Kharzeev}}, \citenamefont
  {{Kikuchi}},\ and\ \citenamefont {{Meyer}}}]{meyer2018}%
  \BibitemOpen
  \bibfield  {author} {\bibinfo {author} {\bibfnamefont {D.~E.}\ \bibnamefont
  {{Kharzeev}}}, \bibinfo {author} {\bibfnamefont {Y.}~\bibnamefont
  {{Kikuchi}}}, \ and\ \bibinfo {author} {\bibfnamefont {R.}~\bibnamefont
  {{Meyer}}},\ }\href {\doibase 10.1140/epjb/e2018-80418-1} {\bibfield
  {journal} {\bibinfo  {journal} {European Physical Journal B}\ }\textbf
  {\bibinfo {volume} {91}},\ \bibinfo {eid} {83} (\bibinfo {year}
  {2018}{\natexlab{b}})},\ \Eprint {http://arxiv.org/abs/1610.08986}
  {arXiv:1610.08986 [cond-mat.mes-hall]} \BibitemShut {NoStop}%
\bibitem [{\citenamefont {{Chang}}\ \emph {et~al.}(2017)\citenamefont
  {{Chang}}, \citenamefont {{Xu}}, \citenamefont {{Wieder}}, \citenamefont
  {{Sanchez}}, \citenamefont {{Huang}}, \citenamefont {{Belopolski}},
  \citenamefont {{Chang}}, \citenamefont {{Zhang}}, \citenamefont {{Bansil}},
  \citenamefont {{Lin}},\ and\ \citenamefont {{Hasan}}}]{chang2017}%
  \BibitemOpen
  \bibfield  {author} {\bibinfo {author} {\bibfnamefont {G.}~\bibnamefont
  {{Chang}}}, \bibinfo {author} {\bibfnamefont {S.-Y.}\ \bibnamefont {{Xu}}},
  \bibinfo {author} {\bibfnamefont {B.~J.}\ \bibnamefont {{Wieder}}}, \bibinfo
  {author} {\bibfnamefont {D.~S.}\ \bibnamefont {{Sanchez}}}, \bibinfo {author}
  {\bibfnamefont {S.-M.}\ \bibnamefont {{Huang}}}, \bibinfo {author}
  {\bibfnamefont {I.}~\bibnamefont {{Belopolski}}}, \bibinfo {author}
  {\bibfnamefont {T.-R.}\ \bibnamefont {{Chang}}}, \bibinfo {author}
  {\bibfnamefont {S.}~\bibnamefont {{Zhang}}}, \bibinfo {author} {\bibfnamefont
  {A.}~\bibnamefont {{Bansil}}}, \bibinfo {author} {\bibfnamefont
  {H.}~\bibnamefont {{Lin}}}, \ and\ \bibinfo {author} {\bibfnamefont {M.~Z.}\
  \bibnamefont {{Hasan}}},\ }\href {\doibase 10.1103/PhysRevLett.119.206401}
  {\bibfield  {journal} {\bibinfo  {journal} {\prl}\ }\textbf {\bibinfo
  {volume} {119}},\ \bibinfo {eid} {206401} (\bibinfo {year}
  {2017})}\BibitemShut {NoStop}%
\bibitem [{\citenamefont {{Rao}}\ \emph {et~al.}(2019)\citenamefont {{Rao}},
  \citenamefont {{Li}}, \citenamefont {{Zhang}}, \citenamefont {{Tian}},
  \citenamefont {{Li}}, \citenamefont {{Fu}}, \citenamefont {{Tang}},
  \citenamefont {{Wang}}, \citenamefont {{Li}}, \citenamefont {{Fan}},
  \citenamefont {{Li}}, \citenamefont {{Huang}}, \citenamefont {{Liu}},
  \citenamefont {{Long}}, \citenamefont {{Fang}}, \citenamefont {{Weng}},
  \citenamefont {{Shi}}, \citenamefont {{Lei}}, \citenamefont {{Sun}},
  \citenamefont {{Qian}},\ and\ \citenamefont {{Ding}}}]{rao2019}%
  \BibitemOpen
  \bibfield  {author} {\bibinfo {author} {\bibfnamefont {Z.}~\bibnamefont
  {{Rao}}}, \bibinfo {author} {\bibfnamefont {H.}~\bibnamefont {{Li}}},
  \bibinfo {author} {\bibfnamefont {T.}~\bibnamefont {{Zhang}}}, \bibinfo
  {author} {\bibfnamefont {S.}~\bibnamefont {{Tian}}}, \bibinfo {author}
  {\bibfnamefont {C.}~\bibnamefont {{Li}}}, \bibinfo {author} {\bibfnamefont
  {B.}~\bibnamefont {{Fu}}}, \bibinfo {author} {\bibfnamefont {C.}~\bibnamefont
  {{Tang}}}, \bibinfo {author} {\bibfnamefont {L.}~\bibnamefont {{Wang}}},
  \bibinfo {author} {\bibfnamefont {Z.}~\bibnamefont {{Li}}}, \bibinfo {author}
  {\bibfnamefont {W.}~\bibnamefont {{Fan}}}, \bibinfo {author} {\bibfnamefont
  {J.}~\bibnamefont {{Li}}}, \bibinfo {author} {\bibfnamefont {Y.}~\bibnamefont
  {{Huang}}}, \bibinfo {author} {\bibfnamefont {Z.}~\bibnamefont {{Liu}}},
  \bibinfo {author} {\bibfnamefont {Y.}~\bibnamefont {{Long}}}, \bibinfo
  {author} {\bibfnamefont {C.}~\bibnamefont {{Fang}}}, \bibinfo {author}
  {\bibfnamefont {H.}~\bibnamefont {{Weng}}}, \bibinfo {author} {\bibfnamefont
  {Y.}~\bibnamefont {{Shi}}}, \bibinfo {author} {\bibfnamefont
  {H.}~\bibnamefont {{Lei}}}, \bibinfo {author} {\bibfnamefont
  {Y.}~\bibnamefont {{Sun}}}, \bibinfo {author} {\bibfnamefont
  {T.}~\bibnamefont {{Qian}}}, \ and\ \bibinfo {author} {\bibfnamefont
  {H.}~\bibnamefont {{Ding}}},\ }\href {\doibase 10.1038/s41586-019-1031-8}
  {\bibfield  {journal} {\bibinfo  {journal} {\nat}\ }\textbf {\bibinfo
  {volume} {567}},\ \bibinfo {pages} {496} (\bibinfo {year}
  {2019})}\BibitemShut {NoStop}%
\bibitem [{\citenamefont {{Rees}}\ \emph {et~al.}(2020)\citenamefont {{Rees}},
  \citenamefont {{Manna}}, \citenamefont {{Lu}}, \citenamefont {{Morimoto}},
  \citenamefont {{Borrmann}}, \citenamefont {{Felser}}, \citenamefont
  {{Moore}}, \citenamefont {{Torchinsky}},\ and\ \citenamefont
  {{Orenstein}}}]{rees2020}%
  \BibitemOpen
  \bibfield  {author} {\bibinfo {author} {\bibfnamefont {D.}~\bibnamefont
  {{Rees}}}, \bibinfo {author} {\bibfnamefont {K.}~\bibnamefont {{Manna}}},
  \bibinfo {author} {\bibfnamefont {B.}~\bibnamefont {{Lu}}}, \bibinfo {author}
  {\bibfnamefont {T.}~\bibnamefont {{Morimoto}}}, \bibinfo {author}
  {\bibfnamefont {H.}~\bibnamefont {{Borrmann}}}, \bibinfo {author}
  {\bibfnamefont {C.}~\bibnamefont {{Felser}}}, \bibinfo {author}
  {\bibfnamefont {J.~E.}\ \bibnamefont {{Moore}}}, \bibinfo {author}
  {\bibfnamefont {D.~H.}\ \bibnamefont {{Torchinsky}}}, \ and\ \bibinfo
  {author} {\bibfnamefont {J.}~\bibnamefont {{Orenstein}}},\ }\href {\doibase
  10.1126/sciadv.aba0509} {\bibfield  {journal} {\bibinfo  {journal} {Science
  Advances}\ }\textbf {\bibinfo {volume} {6}},\ \bibinfo {pages} {eaba0509}
  (\bibinfo {year} {2020})},\ \Eprint {http://arxiv.org/abs/1902.03230}
  {arXiv:1902.03230 [cond-mat.mes-hall]} \BibitemShut {NoStop}%
\bibitem [{\citenamefont {{Ni}}\ \emph {et~al.}(2020)\citenamefont {{Ni}},
  \citenamefont {{Wang}}, \citenamefont {{Zhang}}, \citenamefont {{Pozo}},
  \citenamefont {{Xu}}, \citenamefont {{Han}}, \citenamefont {{Manna}},
  \citenamefont {{Paglione}}, \citenamefont {{Felser}}, \citenamefont
  {{Grushin}}, \citenamefont {{de Juan}}, \citenamefont {{Mele}},\ and\
  \citenamefont {{Wu}}}]{ni2020}%
  \BibitemOpen
  \bibfield  {author} {\bibinfo {author} {\bibfnamefont {Z.}~\bibnamefont
  {{Ni}}}, \bibinfo {author} {\bibfnamefont {K.}~\bibnamefont {{Wang}}},
  \bibinfo {author} {\bibfnamefont {Y.}~\bibnamefont {{Zhang}}}, \bibinfo
  {author} {\bibfnamefont {O.}~\bibnamefont {{Pozo}}}, \bibinfo {author}
  {\bibfnamefont {B.}~\bibnamefont {{Xu}}}, \bibinfo {author} {\bibfnamefont
  {X.}~\bibnamefont {{Han}}}, \bibinfo {author} {\bibfnamefont
  {K.}~\bibnamefont {{Manna}}}, \bibinfo {author} {\bibfnamefont
  {J.}~\bibnamefont {{Paglione}}}, \bibinfo {author} {\bibfnamefont
  {C.}~\bibnamefont {{Felser}}}, \bibinfo {author} {\bibfnamefont {A.~G.}\
  \bibnamefont {{Grushin}}}, \bibinfo {author} {\bibfnamefont {F.}~\bibnamefont
  {{de Juan}}}, \bibinfo {author} {\bibfnamefont {E.~J.}\ \bibnamefont
  {{Mele}}}, \ and\ \bibinfo {author} {\bibfnamefont {L.}~\bibnamefont
  {{Wu}}},\ }\href@noop {} {\  (\bibinfo {year} {2020})},\ \Eprint
  {http://arxiv.org/abs/2006.09612} {arXiv:2006.09612 [cond-mat.mtrl-sci]}
  \BibitemShut {NoStop}%
\bibitem [{\citenamefont {{Ray}}\ \emph {et~al.}(2020)\citenamefont {{Ray}},
  \citenamefont {{Sadhukhan}}, \citenamefont {{Richter}}, \citenamefont
  {{Facio}},\ and\ \citenamefont {{van den Brink}}}]{ray2020}%
  \BibitemOpen
  \bibfield  {author} {\bibinfo {author} {\bibfnamefont {R.}~\bibnamefont
  {{Ray}}}, \bibinfo {author} {\bibfnamefont {B.}~\bibnamefont {{Sadhukhan}}},
  \bibinfo {author} {\bibfnamefont {M.}~\bibnamefont {{Richter}}}, \bibinfo
  {author} {\bibfnamefont {J.~I.}\ \bibnamefont {{Facio}}}, \ and\ \bibinfo
  {author} {\bibfnamefont {J.}~\bibnamefont {{van den Brink}}},\ }\href@noop {}
  {\ ,\ \bibinfo {eid} {arXiv:2006.10602} (\bibinfo {year} {2020})},\ \Eprint
  {http://arxiv.org/abs/2006.10602} {arXiv:2006.10602 [cond-mat.mes-hall]}
  \BibitemShut {NoStop}%
\bibitem [{\citenamefont {Barron}(1986)}]{barron1986}%
  \BibitemOpen
  \bibfield  {author} {\bibinfo {author} {\bibfnamefont {L.~D.}\ \bibnamefont
  {Barron}},\ }\href {\doibase 10.1021/ja00278a029} {\bibfield  {journal}
  {\bibinfo  {journal} {Journal of the American Chemical Society}\ }\textbf
  {\bibinfo {volume} {108}},\ \bibinfo {pages} {5539} (\bibinfo {year}
  {1986})}\BibitemShut {NoStop}%
\bibitem [{\citenamefont {{Soluyanov}}\ \emph {et~al.}(2015)\citenamefont
  {{Soluyanov}}, \citenamefont {{Gresch}}, \citenamefont {{Wang}},
  \citenamefont {{Wu}}, \citenamefont {{Troyer}}, \citenamefont {{Dai}},\ and\
  \citenamefont {{Bernevig}}}]{soluyanov2015}%
  \BibitemOpen
  \bibfield  {author} {\bibinfo {author} {\bibfnamefont {A.~A.}\ \bibnamefont
  {{Soluyanov}}}, \bibinfo {author} {\bibfnamefont {D.}~\bibnamefont
  {{Gresch}}}, \bibinfo {author} {\bibfnamefont {Z.}~\bibnamefont {{Wang}}},
  \bibinfo {author} {\bibfnamefont {Q.}~\bibnamefont {{Wu}}}, \bibinfo {author}
  {\bibfnamefont {M.}~\bibnamefont {{Troyer}}}, \bibinfo {author}
  {\bibfnamefont {X.}~\bibnamefont {{Dai}}}, \ and\ \bibinfo {author}
  {\bibfnamefont {B.~A.}\ \bibnamefont {{Bernevig}}},\ }\href {\doibase
  10.1038/nature15768} {\bibfield  {journal} {\bibinfo  {journal} {\nat}\
  }\textbf {\bibinfo {volume} {527}},\ \bibinfo {pages} {495} (\bibinfo {year}
  {2015})},\ \Eprint {http://arxiv.org/abs/1507.01603} {arXiv:1507.01603
  [cond-mat.mes-hall]} \BibitemShut {NoStop}%
\bibitem [{\citenamefont {{Yang}}\ \emph {et~al.}(2015)\citenamefont {{Yang}},
  \citenamefont {{Liu}}, \citenamefont {{Sun}}, \citenamefont {{Peng}},
  \citenamefont {{Yang}}, \citenamefont {{Zhang}}, \citenamefont {{Zhou}},
  \citenamefont {{Zhang}}, \citenamefont {{Guo}}, \citenamefont {{Rahn}},
  \citenamefont {{Prabhakaran}}, \citenamefont {{Hussain}}, \citenamefont
  {{Mo}}, \citenamefont {{Felser}}, \citenamefont {{Yan}},\ and\ \citenamefont
  {{Chen}}}]{yang2015}%
  \BibitemOpen
  \bibfield  {author} {\bibinfo {author} {\bibfnamefont {L.~X.}\ \bibnamefont
  {{Yang}}}, \bibinfo {author} {\bibfnamefont {Z.~K.}\ \bibnamefont {{Liu}}},
  \bibinfo {author} {\bibfnamefont {Y.}~\bibnamefont {{Sun}}}, \bibinfo
  {author} {\bibfnamefont {H.}~\bibnamefont {{Peng}}}, \bibinfo {author}
  {\bibfnamefont {H.~F.}\ \bibnamefont {{Yang}}}, \bibinfo {author}
  {\bibfnamefont {T.}~\bibnamefont {{Zhang}}}, \bibinfo {author} {\bibfnamefont
  {B.}~\bibnamefont {{Zhou}}}, \bibinfo {author} {\bibfnamefont
  {Y.}~\bibnamefont {{Zhang}}}, \bibinfo {author} {\bibfnamefont {Y.~F.}\
  \bibnamefont {{Guo}}}, \bibinfo {author} {\bibfnamefont {M.}~\bibnamefont
  {{Rahn}}}, \bibinfo {author} {\bibfnamefont {D.}~\bibnamefont
  {{Prabhakaran}}}, \bibinfo {author} {\bibfnamefont {Z.}~\bibnamefont
  {{Hussain}}}, \bibinfo {author} {\bibfnamefont {S.~K.}\ \bibnamefont {{Mo}}},
  \bibinfo {author} {\bibfnamefont {C.}~\bibnamefont {{Felser}}}, \bibinfo
  {author} {\bibfnamefont {B.}~\bibnamefont {{Yan}}}, \ and\ \bibinfo {author}
  {\bibfnamefont {Y.~L.}\ \bibnamefont {{Chen}}},\ }\href {\doibase
  10.1038/nphys3425} {\bibfield  {journal} {\bibinfo  {journal} {Nature
  Physics}\ }\textbf {\bibinfo {volume} {11}},\ \bibinfo {pages} {728}
  (\bibinfo {year} {2015})}\BibitemShut {NoStop}%
\bibitem [{\citenamefont {{Li}}\ \emph {et~al.}(2017)\citenamefont {{Li}},
  \citenamefont {{Wen}}, \citenamefont {{He}}, \citenamefont {{Zhang}},
  \citenamefont {{Xia}}, \citenamefont {{Yu}}, \citenamefont {{Yang}},
  \citenamefont {{Zhu}}, \citenamefont {{Alshareef}},\ and\ \citenamefont
  {{Zhang}}}]{li2017}%
  \BibitemOpen
  \bibfield  {author} {\bibinfo {author} {\bibfnamefont {P.}~\bibnamefont
  {{Li}}}, \bibinfo {author} {\bibfnamefont {Y.}~\bibnamefont {{Wen}}},
  \bibinfo {author} {\bibfnamefont {X.}~\bibnamefont {{He}}}, \bibinfo {author}
  {\bibfnamefont {Q.}~\bibnamefont {{Zhang}}}, \bibinfo {author} {\bibfnamefont
  {C.}~\bibnamefont {{Xia}}}, \bibinfo {author} {\bibfnamefont {Z.-M.}\
  \bibnamefont {{Yu}}}, \bibinfo {author} {\bibfnamefont {S.~A.}\ \bibnamefont
  {{Yang}}}, \bibinfo {author} {\bibfnamefont {Z.}~\bibnamefont {{Zhu}}},
  \bibinfo {author} {\bibfnamefont {H.~N.}\ \bibnamefont {{Alshareef}}}, \ and\
  \bibinfo {author} {\bibfnamefont {X.-X.}\ \bibnamefont {{Zhang}}},\ }\href
  {\doibase 10.1038/s41467-017-02237-1} {\bibfield  {journal} {\bibinfo
  {journal} {Nature Communications}\ }\textbf {\bibinfo {volume} {8}},\
  \bibinfo {eid} {2150} (\bibinfo {year} {2017})}\BibitemShut {NoStop}%
\bibitem [{\citenamefont {Zhang}\ \emph {et~al.}(2018)\citenamefont {Zhang},
  \citenamefont {Yang},\ and\ \citenamefont {Wang}}]{zhang2018}%
  \BibitemOpen
  \bibfield  {author} {\bibinfo {author} {\bibfnamefont {M.}~\bibnamefont
  {Zhang}}, \bibinfo {author} {\bibfnamefont {Z.}~\bibnamefont {Yang}}, \ and\
  \bibinfo {author} {\bibfnamefont {G.}~\bibnamefont {Wang}},\ }\href {\doibase
  10.1021/acs.jpcc.8b00920} {\bibfield  {journal} {\bibinfo  {journal} {The
  Journal of Physical Chemistry C}\ }\textbf {\bibinfo {volume} {122}},\
  \bibinfo {pages} {3533} (\bibinfo {year} {2018})}\BibitemShut {NoStop}%
\bibitem [{\citenamefont {{Xu}}\ \emph {et~al.}(2016)\citenamefont {{Xu}},
  \citenamefont {{Weng}}, \citenamefont {{Lv}}, \citenamefont {{Matt}},
  \citenamefont {{Park}}, \citenamefont {{Bisti}}, \citenamefont {{Strocov}},
  \citenamefont {{Gawryluk}}, \citenamefont {{Pomjakushina}}, \citenamefont
  {{Conder}}, \citenamefont {{Plumb}}, \citenamefont {{Radovic}}, \citenamefont
  {{Aut{\`e}s}}, \citenamefont {{Yazyev}}, \citenamefont {{Fang}},
  \citenamefont {{Dai}}, \citenamefont {{Qian}}, \citenamefont {{Mesot}},
  \citenamefont {{Ding}},\ and\ \citenamefont {{Shi}}}]{xu2016}%
  \BibitemOpen
  \bibfield  {author} {\bibinfo {author} {\bibfnamefont {N.}~\bibnamefont
  {{Xu}}}, \bibinfo {author} {\bibfnamefont {H.~M.}\ \bibnamefont {{Weng}}},
  \bibinfo {author} {\bibfnamefont {B.~Q.}\ \bibnamefont {{Lv}}}, \bibinfo
  {author} {\bibfnamefont {C.~E.}\ \bibnamefont {{Matt}}}, \bibinfo {author}
  {\bibfnamefont {J.}~\bibnamefont {{Park}}}, \bibinfo {author} {\bibfnamefont
  {F.}~\bibnamefont {{Bisti}}}, \bibinfo {author} {\bibfnamefont {V.~N.}\
  \bibnamefont {{Strocov}}}, \bibinfo {author} {\bibfnamefont {D.}~\bibnamefont
  {{Gawryluk}}}, \bibinfo {author} {\bibfnamefont {E.}~\bibnamefont
  {{Pomjakushina}}}, \bibinfo {author} {\bibfnamefont {K.}~\bibnamefont
  {{Conder}}}, \bibinfo {author} {\bibfnamefont {N.~C.}\ \bibnamefont
  {{Plumb}}}, \bibinfo {author} {\bibfnamefont {M.}~\bibnamefont {{Radovic}}},
  \bibinfo {author} {\bibfnamefont {G.}~\bibnamefont {{Aut{\`e}s}}}, \bibinfo
  {author} {\bibfnamefont {O.~V.}\ \bibnamefont {{Yazyev}}}, \bibinfo {author}
  {\bibfnamefont {Z.}~\bibnamefont {{Fang}}}, \bibinfo {author} {\bibfnamefont
  {X.}~\bibnamefont {{Dai}}}, \bibinfo {author} {\bibfnamefont
  {T.}~\bibnamefont {{Qian}}}, \bibinfo {author} {\bibfnamefont
  {J.}~\bibnamefont {{Mesot}}}, \bibinfo {author} {\bibfnamefont
  {H.}~\bibnamefont {{Ding}}}, \ and\ \bibinfo {author} {\bibfnamefont
  {M.}~\bibnamefont {{Shi}}},\ }\href {\doibase 10.1038/ncomms11006} {\bibfield
   {journal} {\bibinfo  {journal} {Nature Communications}\ }\textbf {\bibinfo
  {volume} {7}},\ \bibinfo {eid} {11006} (\bibinfo {year} {2016})},\ \Eprint
  {http://arxiv.org/abs/1507.03983} {arXiv:1507.03983 [cond-mat.mtrl-sci]}
  \BibitemShut {NoStop}%
\bibitem [{\citenamefont {{Modic}}\ \emph {et~al.}(2019)\citenamefont
  {{Modic}}, \citenamefont {{Meng}}, \citenamefont {{Ronning}}, \citenamefont
  {{Bauer}}, \citenamefont {{Moll}},\ and\ \citenamefont
  {{Ramshaw}}}]{modic2019}%
  \BibitemOpen
  \bibfield  {author} {\bibinfo {author} {\bibfnamefont {K.~A.}\ \bibnamefont
  {{Modic}}}, \bibinfo {author} {\bibfnamefont {T.}~\bibnamefont {{Meng}}},
  \bibinfo {author} {\bibfnamefont {F.}~\bibnamefont {{Ronning}}}, \bibinfo
  {author} {\bibfnamefont {E.~D.}\ \bibnamefont {{Bauer}}}, \bibinfo {author}
  {\bibfnamefont {P.~J.~W.}\ \bibnamefont {{Moll}}}, \ and\ \bibinfo {author}
  {\bibfnamefont {B.~J.}\ \bibnamefont {{Ramshaw}}},\ }\href {\doibase
  10.1038/s41598-018-38161-7} {\bibfield  {journal} {\bibinfo  {journal}
  {Scientific Reports}\ }\textbf {\bibinfo {volume} {9}},\ \bibinfo {eid}
  {2095} (\bibinfo {year} {2019})}\BibitemShut {NoStop}%
\bibitem [{\citenamefont {{Yuan}}\ \emph {et~al.}(2020)\citenamefont {{Yuan}},
  \citenamefont {{Zhang}}, \citenamefont {{Zhang}}, \citenamefont {{Yan}},
  \citenamefont {{Lyu}}, \citenamefont {{Zhang}}, \citenamefont {{Li}},
  \citenamefont {{Song}}, \citenamefont {{Zhao}}, \citenamefont {{Leng}},
  \citenamefont {{Ozerov}}, \citenamefont {{Chen}}, \citenamefont {{Wang}},
  \citenamefont {{Shi}}, \citenamefont {{Yan}},\ and\ \citenamefont
  {{Xiu}}}]{yuan2020}%
  \BibitemOpen
  \bibfield  {author} {\bibinfo {author} {\bibfnamefont {X.}~\bibnamefont
  {{Yuan}}}, \bibinfo {author} {\bibfnamefont {C.}~\bibnamefont {{Zhang}}},
  \bibinfo {author} {\bibfnamefont {Y.}~\bibnamefont {{Zhang}}}, \bibinfo
  {author} {\bibfnamefont {Z.}~\bibnamefont {{Yan}}}, \bibinfo {author}
  {\bibfnamefont {T.}~\bibnamefont {{Lyu}}}, \bibinfo {author} {\bibfnamefont
  {M.}~\bibnamefont {{Zhang}}}, \bibinfo {author} {\bibfnamefont
  {Z.}~\bibnamefont {{Li}}}, \bibinfo {author} {\bibfnamefont {C.}~\bibnamefont
  {{Song}}}, \bibinfo {author} {\bibfnamefont {M.}~\bibnamefont {{Zhao}}},
  \bibinfo {author} {\bibfnamefont {P.}~\bibnamefont {{Leng}}}, \bibinfo
  {author} {\bibfnamefont {M.}~\bibnamefont {{Ozerov}}}, \bibinfo {author}
  {\bibfnamefont {X.}~\bibnamefont {{Chen}}}, \bibinfo {author} {\bibfnamefont
  {N.}~\bibnamefont {{Wang}}}, \bibinfo {author} {\bibfnamefont
  {Y.}~\bibnamefont {{Shi}}}, \bibinfo {author} {\bibfnamefont
  {H.}~\bibnamefont {{Yan}}}, \ and\ \bibinfo {author} {\bibfnamefont
  {F.}~\bibnamefont {{Xiu}}},\ }\href {\doibase 10.1038/s41467-020-14749-4}
  {\bibfield  {journal} {\bibinfo  {journal} {Nature Communications}\ }\textbf
  {\bibinfo {volume} {11}},\ \bibinfo {eid} {1259} (\bibinfo {year}
  {2020})}\BibitemShut {NoStop}%
\bibitem [{\citenamefont {{Sun}}\ \emph {et~al.}(2020)\citenamefont {{Sun}},
  \citenamefont {{Song}}, \citenamefont {{Weng}},\ and\ \citenamefont
  {{Dai}}}]{sun2020topological}%
  \BibitemOpen
  \bibfield  {author} {\bibinfo {author} {\bibfnamefont {S.}~\bibnamefont
  {{Sun}}}, \bibinfo {author} {\bibfnamefont {Z.}~\bibnamefont {{Song}}},
  \bibinfo {author} {\bibfnamefont {H.}~\bibnamefont {{Weng}}}, \ and\ \bibinfo
  {author} {\bibfnamefont {X.}~\bibnamefont {{Dai}}},\ }\href {\doibase
  10.1103/PhysRevB.101.125118} {\bibfield  {journal} {\bibinfo  {journal}
  {\prb}\ }\textbf {\bibinfo {volume} {101}},\ \bibinfo {eid} {125118}
  (\bibinfo {year} {2020})},\ \Eprint {http://arxiv.org/abs/1910.01378}
  {arXiv:1910.01378 [cond-mat.mes-hall]} \BibitemShut {NoStop}%
\bibitem [{\citenamefont {{Choi}}\ \emph {et~al.}(2020)\citenamefont {{Choi}},
  \citenamefont {{Villanova}},\ and\ \citenamefont {{Park}}}]{choi2020}%
  \BibitemOpen
  \bibfield  {author} {\bibinfo {author} {\bibfnamefont {Y.}~\bibnamefont
  {{Choi}}}, \bibinfo {author} {\bibfnamefont {J.~W.}\ \bibnamefont
  {{Villanova}}}, \ and\ \bibinfo {author} {\bibfnamefont {K.}~\bibnamefont
  {{Park}}},\ }\href {\doibase 10.1103/PhysRevB.101.035105} {\bibfield
  {journal} {\bibinfo  {journal} {\prb}\ }\textbf {\bibinfo {volume} {101}},\
  \bibinfo {eid} {035105} (\bibinfo {year} {2020})},\ \Eprint
  {http://arxiv.org/abs/1910.04239} {arXiv:1910.04239 [cond-mat.mes-hall]}
  \BibitemShut {NoStop}%
\bibitem [{\citenamefont {{Ruan}}\ \emph {et~al.}(2016)\citenamefont {{Ruan}},
  \citenamefont {{Jian}}, \citenamefont {{Yao}}, \citenamefont {{Zhang}},
  \citenamefont {{Zhang}},\ and\ \citenamefont {{Xing}}}]{ruan2016}%
  \BibitemOpen
  \bibfield  {author} {\bibinfo {author} {\bibfnamefont {J.}~\bibnamefont
  {{Ruan}}}, \bibinfo {author} {\bibfnamefont {S.-K.}\ \bibnamefont {{Jian}}},
  \bibinfo {author} {\bibfnamefont {H.}~\bibnamefont {{Yao}}}, \bibinfo
  {author} {\bibfnamefont {H.}~\bibnamefont {{Zhang}}}, \bibinfo {author}
  {\bibfnamefont {S.-C.}\ \bibnamefont {{Zhang}}}, \ and\ \bibinfo {author}
  {\bibfnamefont {D.}~\bibnamefont {{Xing}}},\ }\href {\doibase
  10.1038/ncomms11136} {\bibfield  {journal} {\bibinfo  {journal} {Nature
  Communications}\ }\textbf {\bibinfo {volume} {7}},\ \bibinfo {eid} {11136}
  (\bibinfo {year} {2016})},\ \Eprint {http://arxiv.org/abs/1511.08284}
  {arXiv:1511.08284 [cond-mat.mes-hall]} \BibitemShut {NoStop}%
\bibitem [{\citenamefont {{Cortijo}}\ \emph {et~al.}(2016)\citenamefont
  {{Cortijo}}, \citenamefont {{Kharzeev}}, \citenamefont {{Landsteiner}},\ and\
  \citenamefont {{Vozmediano}}}]{cortijo2016}%
  \BibitemOpen
  \bibfield  {author} {\bibinfo {author} {\bibfnamefont {A.}~\bibnamefont
  {{Cortijo}}}, \bibinfo {author} {\bibfnamefont {D.}~\bibnamefont
  {{Kharzeev}}}, \bibinfo {author} {\bibfnamefont {K.}~\bibnamefont
  {{Landsteiner}}}, \ and\ \bibinfo {author} {\bibfnamefont {M.~A.~H.}\
  \bibnamefont {{Vozmediano}}},\ }\href {\doibase 10.1103/PhysRevB.94.241405}
  {\bibfield  {journal} {\bibinfo  {journal} {\prb}\ }\textbf {\bibinfo
  {volume} {94}},\ \bibinfo {eid} {241405} (\bibinfo {year} {2016})},\ \Eprint
  {http://arxiv.org/abs/1607.03491} {arXiv:1607.03491 [cond-mat.mes-hall]}
  \BibitemShut {NoStop}%
\bibitem [{\citenamefont {{Shekhar}}\ \emph {et~al.}(2018)\citenamefont
  {{Shekhar}}, \citenamefont {{Kumar}}, \citenamefont {{Grinenko}},
  \citenamefont {{Singh}}, \citenamefont {{Sarkar}}, \citenamefont
  {{Luetkens}}, \citenamefont {{Wu}}, \citenamefont {{Zhang}}, \citenamefont
  {{Komarek}}, \citenamefont {{Kampert}}, \citenamefont {{Skourski}},
  \citenamefont {{Wosnitza}}, \citenamefont {{Schnelle}}, \citenamefont
  {{McCollam}}, \citenamefont {{Zeitler}}, \citenamefont {{K{\"u}bler}},
  \citenamefont {{Yan}}, \citenamefont {{Klauss}}, \citenamefont {{Parkin}},\
  and\ \citenamefont {{Felser}}}]{shekhar2018}%
  \BibitemOpen
  \bibfield  {author} {\bibinfo {author} {\bibfnamefont {C.}~\bibnamefont
  {{Shekhar}}}, \bibinfo {author} {\bibfnamefont {N.}~\bibnamefont {{Kumar}}},
  \bibinfo {author} {\bibfnamefont {V.}~\bibnamefont {{Grinenko}}}, \bibinfo
  {author} {\bibfnamefont {S.}~\bibnamefont {{Singh}}}, \bibinfo {author}
  {\bibfnamefont {R.}~\bibnamefont {{Sarkar}}}, \bibinfo {author}
  {\bibfnamefont {H.}~\bibnamefont {{Luetkens}}}, \bibinfo {author}
  {\bibfnamefont {S.-C.}\ \bibnamefont {{Wu}}}, \bibinfo {author}
  {\bibfnamefont {Y.}~\bibnamefont {{Zhang}}}, \bibinfo {author} {\bibfnamefont
  {A.~C.}\ \bibnamefont {{Komarek}}}, \bibinfo {author} {\bibfnamefont
  {E.}~\bibnamefont {{Kampert}}}, \bibinfo {author} {\bibfnamefont
  {Y.}~\bibnamefont {{Skourski}}}, \bibinfo {author} {\bibfnamefont
  {J.}~\bibnamefont {{Wosnitza}}}, \bibinfo {author} {\bibfnamefont
  {W.}~\bibnamefont {{Schnelle}}}, \bibinfo {author} {\bibfnamefont
  {A.}~\bibnamefont {{McCollam}}}, \bibinfo {author} {\bibfnamefont
  {U.}~\bibnamefont {{Zeitler}}}, \bibinfo {author} {\bibfnamefont
  {J.}~\bibnamefont {{K{\"u}bler}}}, \bibinfo {author} {\bibfnamefont
  {B.}~\bibnamefont {{Yan}}}, \bibinfo {author} {\bibfnamefont {H.~H.}\
  \bibnamefont {{Klauss}}}, \bibinfo {author} {\bibfnamefont {S.~S.~P.}\
  \bibnamefont {{Parkin}}}, \ and\ \bibinfo {author} {\bibfnamefont
  {C.}~\bibnamefont {{Felser}}},\ }\href {\doibase 10.1073/pnas.1810842115}
  {\bibfield  {journal} {\bibinfo  {journal} {Proceedings of the National
  Academy of Science}\ }\textbf {\bibinfo {volume} {115}},\ \bibinfo {pages}
  {9140} (\bibinfo {year} {2018})},\ \Eprint {http://arxiv.org/abs/1604.01641}
  {arXiv:1604.01641 [cond-mat.mtrl-sci]} \BibitemShut {NoStop}%
\bibitem [{\citenamefont {{Ghimire}}\ \emph {et~al.}(2019)\citenamefont
  {{Ghimire}}, \citenamefont {{Facio}}, \citenamefont {{You}}, \citenamefont
  {{Ye}}, \citenamefont {{Checkelsky}}, \citenamefont {{Fang}}, \citenamefont
  {{Kaxiras}}, \citenamefont {{Richter}},\ and\ \citenamefont {{van den
  Brink}}}]{ghimire2019}%
  \BibitemOpen
  \bibfield  {author} {\bibinfo {author} {\bibfnamefont {M.~P.}\ \bibnamefont
  {{Ghimire}}}, \bibinfo {author} {\bibfnamefont {J.~I.}\ \bibnamefont
  {{Facio}}}, \bibinfo {author} {\bibfnamefont {J.-S.}\ \bibnamefont {{You}}},
  \bibinfo {author} {\bibfnamefont {L.}~\bibnamefont {{Ye}}}, \bibinfo {author}
  {\bibfnamefont {J.~G.}\ \bibnamefont {{Checkelsky}}}, \bibinfo {author}
  {\bibfnamefont {S.}~\bibnamefont {{Fang}}}, \bibinfo {author} {\bibfnamefont
  {E.}~\bibnamefont {{Kaxiras}}}, \bibinfo {author} {\bibfnamefont
  {M.}~\bibnamefont {{Richter}}}, \ and\ \bibinfo {author} {\bibfnamefont
  {J.}~\bibnamefont {{van den Brink}}},\ }\href {\doibase
  10.1103/PhysRevResearch.1.032044} {\bibfield  {journal} {\bibinfo  {journal}
  {Physical Review Research}\ }\textbf {\bibinfo {volume} {1}},\ \bibinfo {eid}
  {032044} (\bibinfo {year} {2019})},\ \Eprint
  {http://arxiv.org/abs/1903.03179} {arXiv:1903.03179 [cond-mat.mes-hall]}
  \BibitemShut {NoStop}%
\bibitem [{\citenamefont {{H{\"u}bener}}\ \emph {et~al.}(2017)\citenamefont
  {{H{\"u}bener}}, \citenamefont {{Sentef}}, \citenamefont {{de Giovannini}},
  \citenamefont {{Kemper}},\ and\ \citenamefont {{Rubio}}}]{hubener2017}%
  \BibitemOpen
  \bibfield  {author} {\bibinfo {author} {\bibfnamefont {H.}~\bibnamefont
  {{H{\"u}bener}}}, \bibinfo {author} {\bibfnamefont {M.~A.}\ \bibnamefont
  {{Sentef}}}, \bibinfo {author} {\bibfnamefont {U.}~\bibnamefont {{de
  Giovannini}}}, \bibinfo {author} {\bibfnamefont {A.~F.}\ \bibnamefont
  {{Kemper}}}, \ and\ \bibinfo {author} {\bibfnamefont {A.}~\bibnamefont
  {{Rubio}}},\ }\href {\doibase 10.1038/ncomms13940} {\bibfield  {journal}
  {\bibinfo  {journal} {Nature Communications}\ }\textbf {\bibinfo {volume}
  {8}},\ \bibinfo {eid} {13940} (\bibinfo {year} {2017})},\ \Eprint
  {http://arxiv.org/abs/1604.03399} {arXiv:1604.03399 [cond-mat.mtrl-sci]}
  \BibitemShut {NoStop}%
\bibitem [{\citenamefont {{He}}\ \emph {et~al.}(2018)\citenamefont {{He}},
  \citenamefont {{Di Sante}}, \citenamefont {{Li}}, \citenamefont {{Chen}},
  \citenamefont {{Rondinelli}},\ and\ \citenamefont {{Franchini}}}]{he2018}%
  \BibitemOpen
  \bibfield  {author} {\bibinfo {author} {\bibfnamefont {J.}~\bibnamefont
  {{He}}}, \bibinfo {author} {\bibfnamefont {D.}~\bibnamefont {{Di Sante}}},
  \bibinfo {author} {\bibfnamefont {R.}~\bibnamefont {{Li}}}, \bibinfo {author}
  {\bibfnamefont {X.-Q.}\ \bibnamefont {{Chen}}}, \bibinfo {author}
  {\bibfnamefont {J.~M.}\ \bibnamefont {{Rondinelli}}}, \ and\ \bibinfo
  {author} {\bibfnamefont {C.}~\bibnamefont {{Franchini}}},\ }\href {\doibase
  10.1038/s41467-017-02814-4} {\bibfield  {journal} {\bibinfo  {journal}
  {Nature Communications}\ }\textbf {\bibinfo {volume} {9}},\ \bibinfo {eid}
  {492} (\bibinfo {year} {2018})}\BibitemShut {NoStop}%
\bibitem [{\citenamefont {{Sharma}}\ \emph {et~al.}(2019)\citenamefont
  {{Sharma}}, \citenamefont {{Xiang}}, \citenamefont {{Shao}}, \citenamefont
  {{Zhang}}, \citenamefont {{Tsymbal}}, \citenamefont {{Hamilton}},\ and\
  \citenamefont {{Seidel}}}]{sharma2019}%
  \BibitemOpen
  \bibfield  {author} {\bibinfo {author} {\bibfnamefont {P.}~\bibnamefont
  {{Sharma}}}, \bibinfo {author} {\bibfnamefont {F.-X.}\ \bibnamefont
  {{Xiang}}}, \bibinfo {author} {\bibfnamefont {D.-F.}\ \bibnamefont {{Shao}}},
  \bibinfo {author} {\bibfnamefont {D.}~\bibnamefont {{Zhang}}}, \bibinfo
  {author} {\bibfnamefont {E.~Y.}\ \bibnamefont {{Tsymbal}}}, \bibinfo {author}
  {\bibfnamefont {A.~R.}\ \bibnamefont {{Hamilton}}}, \ and\ \bibinfo {author}
  {\bibfnamefont {J.}~\bibnamefont {{Seidel}}},\ }\href {\doibase
  10.1126/sciadv.aax5080} {\bibfield  {journal} {\bibinfo  {journal} {Science
  Advances}\ }\textbf {\bibinfo {volume} {5}},\ \bibinfo {eid} {eaax5080}
  (\bibinfo {year} {2019})}\BibitemShut {NoStop}%
\bibitem [{\citenamefont {{O'Brien}}\ \emph {et~al.}(2017)\citenamefont
  {{O'Brien}}, \citenamefont {{Beenakker}},\ and\ \citenamefont
  {{Adagideli}}}]{obrien2017}%
  \BibitemOpen
  \bibfield  {author} {\bibinfo {author} {\bibfnamefont {T.~E.}\ \bibnamefont
  {{O'Brien}}}, \bibinfo {author} {\bibfnamefont {C.~W.~J.}\ \bibnamefont
  {{Beenakker}}}, \ and\ \bibinfo {author} {\bibfnamefont {I.}~\bibnamefont
  {{Adagideli}}},\ }\href {\doibase 10.1103/PhysRevLett.118.207701} {\bibfield
  {journal} {\bibinfo  {journal} {\prl}\ }\textbf {\bibinfo {volume} {118}},\
  \bibinfo {eid} {207701} (\bibinfo {year} {2017})},\ \Eprint
  {http://arxiv.org/abs/1612.06848} {arXiv:1612.06848 [cond-mat.mes-hall]}
  \BibitemShut {NoStop}%
\bibitem [{\citenamefont {{Hirschberger}}\ \emph {et~al.}(2016)\citenamefont
  {{Hirschberger}}, \citenamefont {{Kushwaha}}, \citenamefont {{Wang}},
  \citenamefont {{Gibson}}, \citenamefont {{Liang}}, \citenamefont {{Belvin}},
  \citenamefont {{Bernevig}}, \citenamefont {{Cava}},\ and\ \citenamefont
  {{Ong}}}]{hirschberger2016}%
  \BibitemOpen
  \bibfield  {author} {\bibinfo {author} {\bibfnamefont {M.}~\bibnamefont
  {{Hirschberger}}}, \bibinfo {author} {\bibfnamefont {S.}~\bibnamefont
  {{Kushwaha}}}, \bibinfo {author} {\bibfnamefont {Z.}~\bibnamefont {{Wang}}},
  \bibinfo {author} {\bibfnamefont {Q.}~\bibnamefont {{Gibson}}}, \bibinfo
  {author} {\bibfnamefont {S.}~\bibnamefont {{Liang}}}, \bibinfo {author}
  {\bibfnamefont {C.~A.}\ \bibnamefont {{Belvin}}}, \bibinfo {author}
  {\bibfnamefont {B.~A.}\ \bibnamefont {{Bernevig}}}, \bibinfo {author}
  {\bibfnamefont {R.~J.}\ \bibnamefont {{Cava}}}, \ and\ \bibinfo {author}
  {\bibfnamefont {N.~P.}\ \bibnamefont {{Ong}}},\ }\href {\doibase
  10.1038/nmat4684} {\bibfield  {journal} {\bibinfo  {journal} {Nature
  Materials}\ }\textbf {\bibinfo {volume} {15}},\ \bibinfo {pages} {1161}
  (\bibinfo {year} {2016})},\ \Eprint {http://arxiv.org/abs/1602.07219}
  {arXiv:1602.07219 [cond-mat.str-el]} \BibitemShut {NoStop}%
\bibitem [{\citenamefont {{Cano}}\ \emph {et~al.}(2017)\citenamefont {{Cano}},
  \citenamefont {{Bradlyn}}, \citenamefont {{Wang}}, \citenamefont
  {{Hirschberger}}, \citenamefont {{Ong}},\ and\ \citenamefont
  {{Bernevig}}}]{cano2017}%
  \BibitemOpen
  \bibfield  {author} {\bibinfo {author} {\bibfnamefont {J.}~\bibnamefont
  {{Cano}}}, \bibinfo {author} {\bibfnamefont {B.}~\bibnamefont {{Bradlyn}}},
  \bibinfo {author} {\bibfnamefont {Z.}~\bibnamefont {{Wang}}}, \bibinfo
  {author} {\bibfnamefont {M.}~\bibnamefont {{Hirschberger}}}, \bibinfo
  {author} {\bibfnamefont {N.~P.}\ \bibnamefont {{Ong}}}, \ and\ \bibinfo
  {author} {\bibfnamefont {B.~A.}\ \bibnamefont {{Bernevig}}},\ }\href
  {\doibase 10.1103/PhysRevB.95.161306} {\bibfield  {journal} {\bibinfo
  {journal} {\prb}\ }\textbf {\bibinfo {volume} {95}},\ \bibinfo {eid} {161306}
  (\bibinfo {year} {2017})},\ \Eprint {http://arxiv.org/abs/1604.08601}
  {arXiv:1604.08601 [cond-mat.mes-hall]} \BibitemShut {NoStop}%
\bibitem [{\citenamefont {{Qu}}\ \emph {et~al.}(2016)\citenamefont {{Qu}},
  \citenamefont {{van Veen}}, \citenamefont {{de Vries}}, \citenamefont
  {{Beukman}}, \citenamefont {{Wimmer}}, \citenamefont {{Yi}}, \citenamefont
  {{Kiselev}}, \citenamefont {{Nguyen}}, \citenamefont {{Sokolich}},
  \citenamefont {{Manfra}}, \citenamefont {{Nichele}}, \citenamefont
  {{Marcus}},\ and\ \citenamefont {{Kouwenhoven}}}]{qu2016}%
  \BibitemOpen
  \bibfield  {author} {\bibinfo {author} {\bibfnamefont {F.}~\bibnamefont
  {{Qu}}}, \bibinfo {author} {\bibfnamefont {J.}~\bibnamefont {{van Veen}}},
  \bibinfo {author} {\bibfnamefont {F.~K.}\ \bibnamefont {{de Vries}}},
  \bibinfo {author} {\bibfnamefont {A.~J.~A.}\ \bibnamefont {{Beukman}}},
  \bibinfo {author} {\bibfnamefont {M.}~\bibnamefont {{Wimmer}}}, \bibinfo
  {author} {\bibfnamefont {W.}~\bibnamefont {{Yi}}}, \bibinfo {author}
  {\bibfnamefont {A.~A.}\ \bibnamefont {{Kiselev}}}, \bibinfo {author}
  {\bibfnamefont {B.-M.}\ \bibnamefont {{Nguyen}}}, \bibinfo {author}
  {\bibfnamefont {M.}~\bibnamefont {{Sokolich}}}, \bibinfo {author}
  {\bibfnamefont {M.~J.}\ \bibnamefont {{Manfra}}}, \bibinfo {author}
  {\bibfnamefont {F.}~\bibnamefont {{Nichele}}}, \bibinfo {author}
  {\bibfnamefont {C.~M.}\ \bibnamefont {{Marcus}}}, \ and\ \bibinfo {author}
  {\bibfnamefont {L.~P.}\ \bibnamefont {{Kouwenhoven}}},\ }\href {\doibase
  10.1021/acs.nanolett.6b03297} {\bibfield  {journal} {\bibinfo  {journal}
  {Nano Letters}\ }\textbf {\bibinfo {volume} {16}},\ \bibinfo {pages} {7509}
  (\bibinfo {year} {2016})},\ \Eprint {http://arxiv.org/abs/1608.05478}
  {arXiv:1608.05478 [cond-mat.mes-hall]} \BibitemShut {NoStop}%
\bibitem [{\citenamefont {{Nilsson}}\ \emph {et~al.}(2009)\citenamefont
  {{Nilsson}}, \citenamefont {{Caroff}}, \citenamefont {{Thelander}},
  \citenamefont {{Larsson}}, \citenamefont {{Wagner}}, \citenamefont
  {{Wernersson}}, \citenamefont {{Samuelson}},\ and\ \citenamefont
  {{Xu}}}]{nilsson2009}%
  \BibitemOpen
  \bibfield  {author} {\bibinfo {author} {\bibfnamefont {H.~A.}\ \bibnamefont
  {{Nilsson}}}, \bibinfo {author} {\bibfnamefont {P.}~\bibnamefont {{Caroff}}},
  \bibinfo {author} {\bibfnamefont {C.}~\bibnamefont {{Thelander}}}, \bibinfo
  {author} {\bibfnamefont {M.}~\bibnamefont {{Larsson}}}, \bibinfo {author}
  {\bibfnamefont {J.~B.}\ \bibnamefont {{Wagner}}}, \bibinfo {author}
  {\bibfnamefont {L.-E.}\ \bibnamefont {{Wernersson}}}, \bibinfo {author}
  {\bibfnamefont {L.}~\bibnamefont {{Samuelson}}}, \ and\ \bibinfo {author}
  {\bibfnamefont {H.~Q.}\ \bibnamefont {{Xu}}},\ }\href {\doibase
  10.1021/nl901333a} {\bibfield  {journal} {\bibinfo  {journal} {Nano Letters}\
  }\textbf {\bibinfo {volume} {9}},\ \bibinfo {pages} {3151} (\bibinfo {year}
  {2009})}\BibitemShut {NoStop}%
\bibitem [{\citenamefont {{Vurgaftman}}\ \emph {et~al.}(2001)\citenamefont
  {{Vurgaftman}}, \citenamefont {{Meyer}},\ and\ \citenamefont
  {{Ram-Mohan}}}]{vurgaftman2001}%
  \BibitemOpen
  \bibfield  {author} {\bibinfo {author} {\bibfnamefont {I.}~\bibnamefont
  {{Vurgaftman}}}, \bibinfo {author} {\bibfnamefont {J.~R.}\ \bibnamefont
  {{Meyer}}}, \ and\ \bibinfo {author} {\bibfnamefont {L.~R.}\ \bibnamefont
  {{Ram-Mohan}}},\ }\href {\doibase 10.1063/1.1368156} {\bibfield  {journal}
  {\bibinfo  {journal} {Journal of Applied Physics}\ }\textbf {\bibinfo
  {volume} {89}},\ \bibinfo {pages} {5815} (\bibinfo {year}
  {2001})}\BibitemShut {NoStop}%
\end{thebibliography}%

\end{document}